\documentclass{article}

\usepackage{PRIMEarxiv}

\usepackage[utf8]{inputenc} 
\usepackage[T1]{fontenc}    
\usepackage{hyperref}       
\usepackage{url}            
\usepackage{booktabs}       
\usepackage{amsfonts}       
\usepackage{nicefrac}       
\usepackage{microtype}      
\usepackage{lipsum}
\usepackage{amsmath,amssymb,amsfonts}
\usepackage{stfloats}
\usepackage{fancyhdr}       
\usepackage{graphicx}       
\graphicspath{{media/}}     

\usepackage{subfig}
\newcommand*{\medcup}{\mathbin{\scalebox{2.2}{\ensuremath{\cup}}}}

\begin{document}
  
\title{Secure Event-Triggered Control for Vehicle Platooning in the Presence of Modification Attacks }

\author{Ali Nikoutadbir\\
	Dept. of Electrical and Computer Engineering\\
	Tarbiat Modares University\\
	Tehran, Iran \\
	\texttt{alinikotadbir@modares.ac.ir} \\
	\And
	Sajjad Torabi \\
	Dept. of Electrical and Computer Engineering\\
	Tarbiat Modares University\\
	Tehran, Iran \\
	\texttt{storabi@modares.ac.ir} \\
    \And
	Sadegh Bolouki \\
	Dept. of Mechanical Engineering\\
	Polytechnique Montréal\\
	Montreal, Quebec, Canada \\
	\texttt{sadegh.bolouki@polymtl.ca} \\
}

\maketitle

\begin{abstract}
This paper addresses the problem of achieving secure consensus in a vehicular platoon using event-triggered control. The platoon consists of a leader and multiple follower vehicles exchanging their position and velocity information discretely to maintain stability. The paper focuses on the issue of gain modification attacks, where a malicious actor attempts to alter the controller gains to destabilize the platoon. To tackle this problem, the paper proposes a resilient event-triggered control scheme that ensures secure consensus while considering constraints on the duration and frequency of attacks. The paper also introduces an attack mitigation strategy through a topology-switching procedure, which is employed when the attack frequency and duration constraints are not met. First, sufficient consensus conditions for distributed static and dynamic event-triggered control schemes are derived under sequential gain modification attacks. The impact of the system matrices and triggering parameters on the attack constraints is also discussed. Second, the paper derives conditions to broaden the range of controller gain stability using the Schur stability criterion to mitigate attacks and maintain platoon stability. Finally, the effectiveness of the proposed methodology is demonstrated through simulation scenarios in various case studies.
\end{abstract}


\section{Introduction}
with increasing population, economic activities, and mobility, the demand for transportation has increased over the past few decades. The vehicular platoon has the advantage of increasing safety, improving traffic capacity, and reducing fuel consumption, which acts as one of the attractive problems in the field of autonomous vehicles and Intelligent Transportation systems \cite{chen2010review,caveney2010cooperative,zheng2015stability}. A vehicular platoon, with the cooperative control of multiple vehicles, is a group of vehicles moving at the same speed while maintaining the desired distance. As a multiagent system (MAS), platoon vehicles follow an information flow topology called a communication network for interaction. Through this network, vehicles exchange their information including position, velocity, and acceleration, to their neighboring vehicles. The effects of agent dynamics and network topology on the consensus of linear discrete-time leaderless and leader-following MASs were investigated in \cite{you2011network,yan2020analysis}. One of the main challenges in vehicular platoon control is the limited communication resources and energy constraints. The majority of the existing communication imperfections (such as communication time delays, packet dropout, channel fading, etc.) only appear when the communication channels are widely used by many vehicles \cite{wang2020adaptive}.
Recently, much research has been done to design efficient control strategies that can prevent excessive use of communication resources. One of the known solutions to reduce communication and the computational burden is time-triggered control, but this wastes the communication resources due to information broadcasting at every sampling instant, especially when the sampling intervals are very short \cite{zhang2020distributeds,nowzari2019event, Tabuada2007real-time}. To abandon the paradigm of periodic sampling, the event-triggered control (ETC) scheme as an effective control system has been introduced. This control scheme can decrease the unnecessary waste of communication resources as it can sample and update the states of the controllers only when the local measurement error exceeds a certain threshold \cite{ding2017overview,liu2020event,dimarogonas2009event}. Zheng-Guang Wu et al.\cite{wu2017event} developed the consensus of MASs via a novel ETC designed based on the estimated state of neighboring agents for fixed and switching topologies. The leader-following consensus of MASs via the ETC scheme in distributed, centralized, and clustered settings for various network topologies was investigated in \cite{xu2015event}. The event-triggered control has been used in platoon systems \cite{liu2019cooperative,linsenmayer2015event,li2019string,wen2018event,ge2021dynamic}. A distributed event-triggered communication strategy for cooperative platoon control of heterogeneous vehicles \cite{liu2019cooperative}, and the ETC of platoons via actuator delays was proposed in \cite{wen2018event}.
Generally, when the ETC system is close to reaching a consensus, the relative error tends to be close to zero. Then, there are instants where the system may be triggered unnecessarily. To address this issue, dynamic event-triggered control (DETC) is employed to extend the capabilities of ETC and handle such situations which has recently attracted interest in vehicle platooning applications. In DETC, an additional internal variable is used to dynamically regulate the threshold level for each agent which in effect makes the threshold smaller over time and closer to the relative error \cite{ge2021dynamic,ge2020dynamic,he2019adaptive,xiao2021dynamic}. For instance, the DETC of platoons over-resource-constrained VANET was examined \cite{ge2021dynamic} and \cite{ge2020dynamic} conducted the recent progress regarding its motivation, techniques, and challenges for DETC in platoons. In \cite{xiao2021dynamic}, the DETC of automated vehicles under random communication topologies was investigated. Note that discrete-time systems are more prevalent due to their ease of implementation for practical applications in networked control systems \cite{karaki2021distributed}. Mishra et al. \cite{mishra2021dynamic,mishra2021average} propose static and dynamic event-triggered control schemes for multi-agent systems with linear dynamics in the discrete-time domain, which inherently avoids the Zeno phenomenon.
Security of cyber-physical systems, such as vehicle platoons, is among the critical aspects of ensuring their suitable operation and reception by society. Absolutely, the control structure of vehicle platoons is susceptible to various attacks that can lead to oscillation, accidents, and destabilization. Implementing effective security measures is crucial for establishing a safe and secure consensus. Resilience against cyber-attacks is a critical concern for vehicular platoons, as attackers can exploit vulnerabilities in communication and control systems \cite{ju2020deception,zhang2020distributed,mousavinejad2019distributed,petrillo2020secure,dadras2015vehicular,dadras2019resilient,khanapuri2021learning}. The former refers to study attacks on communication such as \cite{ju2020deception}, where proposed a framework for detecting deception attacks on vehicular platoons and in \cite{zhang2020distributed} proposed a switched time-delay system approach for distributed secure platoon control problems with Denial-of-Service (DoS) attacks. In \cite{mousavinejad2019distributed}, a designed filter containing two ellipsoidal sets is used to detect cyber-attacks, and a recovery mechanism is provided. The latter refers to the attacks on control systems where an adversary alters a subset of control inputs, sensor measurements, or control laws such as gain modification attacks. In \cite{dadras2015vehicular}, demonstrated that gain modification attacks can lead to collisions and destabilization by locally manipulating the control law. To address this issue, they proposed a fractional-order control scheme to prevent collisions in a hostile platooning environment \cite{dadras2019resilient}. Khanapuri et al. \cite{khanapuri2021learning} represented that changing the gain values of unattacked vehicles within a platoon can help mitigate an attack.
Besides the type of attack with respect to the target component, how the adversary launches the attack is a matter of importance. A sequential attack is a commonly used model, where the attacker's activity is characterized by average attack frequency and duration limits. Secure consensus in systems, it has been demonstrated that exponential secure consensus can be achieved as long as the attack duration and frequency remain below specific critical values \cite{xu2019event,de2015input,feng2019secure,he2021secure,liang2021secure}. Xu et al. \cite{xu2019event}, and Feng et al. \cite{feng2019secure} explored centralized and distributed event-triggered secure consensus of general linear MASs subject to sequential DoS attacks, respectively. The secure consensus of linear MASs under event-triggered control suffers from a sequential scaling attack has been considered in \cite{he2021secure}. In \cite{xiao2020resilient}, the problem of resilient distributed event-triggered platoon control under energy-limited DoS attacks was studied. Moreover, a novel event-triggered control problem was proposed under sequential DoS and deception attacks for a platoon of vehicles \cite{xu2022event}. Designing resilient distributed ETC and DETC against malicious attacks is in great demand.
Although most of the existing results on sequential DoS and deception attacks pertain to those directed toward the communication infrastructure, sequential attacks on the control signal have rarely been considered, especially for vehicle platooning via event-triggered control. In this paper, we consider a scenario where an adversary manipulates the control signal for a limited time and then remains dormant to conserve energy for future attacks. Unlike sequential DoS attacks, where the system lacks control over attack intervals and truly affected intervals are determined by a time-triggered detection mechanism \cite{xu2019event,feng2019secure}, our study similar to \cite{he2021secure}, remains under control despite tampered signals. Since the control inputs can only be updated at triggering times, the attacker cannot affect the control law effect within two consecutive triggering times. This implies that the adverse effect is influenced by both the event-triggering sequence and the properties of the attack, posing challenges for analysis. This paper addresses the description of the affected interval and safe interval within the event-triggered framework, focusing on the secure consensus of platoons subject to sequential gain modification attacks through event-triggered control. Additionally, we propose a topology-switching scheme to enhance system resilience against attacks without specific constraints on attack frequency and duration. Our motivation is to develop a discrete-time control methodology that can effectively combat control input cyber-attacks while optimizing the utilization of communication resources within platoons.

This paper presents two main contributions of secure consensus in vehicular platoons under gain modification attacks. Firstly, we investigate the secure consensus of a general linear discrete-time platoon system using static/dynamic event-triggered schemes. Our focus is on the impact of sequential gain modification attacks on the system's performance. Secondly, we propose a novel topology-switching approach to mitigate the attacks and stabilize the platoon. This approach broadens the range of the controller's stability by increasing the allowable bound of control gains, achieved through the reduction of the largest eigenvalue of the system. Numerical examples are provided to validate the theoretical results and demonstrate the effectiveness of the proposed methodology.

The rest of this paper is organized as follows. Section II provides the Problem formulation and preliminaries of the basic graphs theory and attack model. Section III presents the distributed static and dynamic event-triggered control scheme. Section IV explains the attack mitigation results and consensus criteria are derived. The simulation results are provided in Section V. Finally, some conclusions are drawn in Section VI.

Notation:
The set of n-dimensional real column vectors and dimensional real matrices is expressed as ${{\mathbb{R}}^{n}}$ and ${{\mathbb{R}}^{n\times m}}$, respectively. For a real symmetric matrix $A$, $A^T$ is the transpose of the matrix $A$. $\lambda_{max}(A)$ and $\lambda_{min}(A)$ denote the largest and the smallest eigenvalues of $A$, respectively. $\sigma_{min}(A)$ and $\sigma_{max}(A)$ denote the smallest and largest singular values of $A$, respectively.The symbol $\otimes$ is used to represent the matrix Kronecker product and $\left\Vert{ . }\right\Vert$ denote the Euclidean vector norm.

\section{Problem formulation and preliminaries}
\label{sec:problem_formulation}

\subsection{Graph Theory}
Consider a platoon with $N+1$ vehicles, consisting of one leader and $N$ following vehicles, which are labeled from the leader to the tail by $0,1,\ldots,N$, where vehicle 0 is the leader vehicle. An undirected graph denoted as $G=(V,E,A)$ is utilized to describe the communication topology amongst the following vehicles, where the node set $V=\left\{1,2,\ldots, N \right\}$ corresponds to the finite set of vehicles and $E \subset V \times V$ denotes the set of edges, representing the communication channels between pairs of vehicles.  Let $\mathcal{N}_i \,=\, \{ j\in V \,|\, (j,i)\in E \} $ represent the neighbor set of the vehicle $i$. The graph structure is described by the adjacency matrix $A=[a_{ij}]\in \mathbb{R}^{N\times N}$, where $a_{ij}=1$ if $(i,j) \in E$, which means vehicle $i$ receives information from vehicle $j$, and $a_{ij}=0$ otherwise. $G$ is called undirected if $ (i,j) \in E ~\Leftrightarrow~ (j,i) \in E,$
which means that the communication channels are bidirectional. The Laplacian matrix of graph $G$ is defined as $L=\left( {{l}_{ij}} \right)\in {\mathbb{R}^{N\times N}}$, where ${{l}_{ij}}=-{{a}_{ij}}$ for $j\ne i$, and ${{l}_{ii}}=\sum_{j\in \mathcal{N}_{i}}{{a}_{ij}}$. Then, we define a diagonal pinning matrix $D=diag\left\{ {{d}_{1}},{{d}_{2}},\ldots,{{d}_{N}} \right\}$ where ${{d}_{i}}$ = 1 if the vehicle $i$ can receive the state information directly from the leader vehicle and ${{d}_{i}} = 0$ otherwise. The platoon with a leader can be described by a graph $\tilde{G}=\left(\tilde{V},\tilde{E},\tilde{A} \right)$ that contains $G$ as a subgraph, where $\tilde{V} =\left\{0,1,2,\ldots,N \right\}$ and an edge subset $\tilde{E} \subset \tilde{V} \times \tilde{V}$. The graph $\tilde{G}$ is called connected if at least a path exists from the leader vehicle to every other following vehicle. Moreover, define the matrix $H=L+D$, it should be clear that $H$ is symmetric. Matrix $H$ is positive definite, if the graph $\tilde{G}$ is connected, and its eigenvalues can be set in ascending order: $0<{{\lambda }_{1}}\le{{\lambda }_{2}}\le\ldots \le{{\lambda }_{N}}$.

\subsection{Problem Formulation}

The vehicular platoon travels along a straight road, and all the members exchange information with each other. Generally, the absolute position and the velocity are transmitted through a communication network. The discrete-time dynamics of a follower vehicle $i$, $1 \leq i \leq N$, can be described as
\begin{equation}
{{x}_{i}}\left( t+1 \right)=A{{x}_{i}}\left( t \right)+\text{B}{{u}_{i}}\left( t \right)
\label{eq_1}
\end{equation}
where ${{x}_{i}}\left( t \right)=\left[ \begin{matrix}
   {{p}_{i}}\left( t \right)  & {{v}_{i}}\left( t\right)
\end{matrix} \right]^T \in {{\mathbb{R}}^{2}}$ is the state of the vehicle $i$ and ${{p}_{i}}\left( t \right)$, ${{v}_{i}}\left( t \right)$ represent its position and velocity, respectively, and ${{u}_{i}}\in {{\mathbb{R}}^{1}}$ denotes its control input. Furthermore,$A= \begin{bmatrix}
   1 & \hat{T}  \\
   0 & 1 \\
\end{bmatrix}  \in {{\mathbb{R}}^{2\times 2}},~B=\left[ \begin{matrix}
   0 \\
\hat{T} \\
\end{matrix} \right] \in {{\mathbb{R}}^{2\times 1}} $ are constant matrices and $\hat{T}$ is the sampling time. It shows the pair (A, B) is stabilizable. The dynamics of the leader vehicle are described as
\begin{equation}
{x_0}\left( t+1 \right)=A{{x}_{0}}\left( t \right)
\label{eq_2}
\end{equation}
where ${{x}_{0}}\left( t \right)$ is the state of the leader.

\emph{Assumption 1:}
    The graph $G$ is undirected and connected.   
    \label{assumption:1}


 The distributed  event-triggered controller is developed  as
\begin{equation}
{{u}_{i}}\left( t \right)=K~{{\hat{q}}_{i}}\left( t_{k}^{i} \right),
\label{eq_3}
\end{equation}
for $t\in \left[ t_{k}^{i}~,t_{k+1}^{i} \right)$. Also, the combined measurement is defined as
\begin{equation}
{{\hat{q}}_{i}}\left( t \right)=\underset{j\in {{N}_{i}}}{\mathop \sum }\,\left( {{{\hat{x}}}_{j}}\left( t \right)-{{{\hat{x}}}_{i}}\left( t \right) \right)+{{m}_{i}}\left( {{x}_{0}}\left( t \right)-{{{\hat{x}}}_{i}}\left( t \right) \right)
\label{eq_4}
\end{equation}

where $K=\left[ {{k}_{p}},{{k}_{v}} \right]$, $K\in {{\mathbb{R}}^{1\times 2}}$ is the gain matrix of the controller to be determined later. Also, ${{m}_{i}}$ is supposed as to be 1 if vehicle $i$ can get information from the leader and ${{m}_{i}}$ = 0 otherwise. Let $t_{0}^{i}, t_{1}^{i}, \ldots, t_{k}^{i}$ represent the sequence of event times of vehicle $i$ such that $t_{k}^{i}$ is the ${k_th}$ triggering instant. Then, triggering instants are determined only when a local triggering condition is violated. The recipients maintain the last broadcast state of vehicle $i$ at time step $t$ as an estimate, denoted by ${{\hat{x}}_{i}}\left( t \right)$. More specifically,

\begin{equation}
\hat{x}_i(t) = x_i(t_{k}^{i}),\quad t\in [t_{k}^{i}, t_{k+1}^{i}),\quad j\in \{ i \}\cup \mathcal{N}_{i} 
\label{eq_5}
\end{equation}
\begin{equation}
t_{k}^{i} =
\begin{cases}
  t_{s}, & \text{if topology is switched} \\
  t_{E}, & \text{event-triggered instants}
\end{cases}
\label{eq_6}
\end{equation}

Vehicle $i$ only has access to the latest broadcasted states of its neighbors $\hat{x}_j(t)$ instead of their true ones $x_j(t)$. We use the estimated state $\hat{x}_j$ rather than the real state $x_j$. It is worth noting that $\hat{x}_0(t)$ is replaced by $x_0(t)$ in equation (\ref{eq_4}).

Since there exists no control input for the leader. Based on equation (\ref{eq_6}),
the information exchange is indeed necessary in two cases. Firstly, during a topology switch at time $t=t_{s}$, this is necessary to ensure that the vehicles can establish communication and update their states based on the new network topology. Secondly, information exchange is also required when vehicles satisfy the event-triggering condition at time $t={{t}_{E}}$. This exchange of information allows the vehicles to synchronize and maintain consensus within the platoon. As a consequence of this setting, the control inputs between the triggering times remain constant.

The objective of the consensus platoon problem is to ensure that the corresponding elements of each vehicle's state converge to a single trajectory. This goal can be expressed formally as follows :
\begin{equation}
\underset{t\overset{~}{\mathop{\to }}\,\infty }{\mathop{\lim }}\,\left| {{v}_{i}}\left( t \right)-{{v}_{0}}\left( t \right) \right|=0
\label{eq_7}
\end{equation}
\begin{equation}
\underset{t\overset{~}{\mathop{\to }}\,\infty }{\mathop{\lim }}\,\left| {{p}_{i}}\left( t \right)-{{p}_{0}}\left( t \right)-{{\nabla }_{i,0}} \right|=0
\label{eq_8}
\end{equation}
where ${{\nabla }_{i,0}}$ is the desired position between the vehicle $i$ and the leader vehicle $0$.

Similar to \cite{mishra2021dynamic}, it is known that in the nominal case when the original state and the broadcast state match at all times as ${{\hat{x}}_{i}}\left( t \right)={{x}_{i}}\left( t \right)$ for $i\in \text{V}$.
For a leader-follower discrete-time system, a necessary and sufficient condition for consensus is the existence of a feedback gain matrix $K$ such that the following spectral radius is lower than one:
\begin{equation}
J=diag\left( A-{{\lambda }_{1}}BK,\ldots ,A-{{\lambda }_{N}}BK \right), ~~\rho \left( J \right)<1
\label{eq_9}
\end{equation}
We assume that such a matrix $K$ exists. To specify the event instant, we introduce the measurement error of the vehicle $i$ as
\begin{equation}
{{e}_{i}}\left( t \right)=~{{\hat{x}}_{i}}\left( t \right)-~{{x}_{i}}\left( t \right)
\label{eq_10}
\end{equation}
Define ${{\delta }_{i}}={{x}_{i}}-{{x}_{0}}$, which represents the difference between the states of the vehicle $i$ and the leader. Let
\begin{equation}
{x}\left( t \right)={{\left[ ~{{x}_{1}}\left( \text{t} \right) ,\ldots,{{x}_{N}}\left( \text{t} \right) \right]}^{T}}
\nonumber\end{equation}
\begin{equation}
\hat{x}\left( t \right)={{\left[ ~{{{\hat{x}}}_{1}}\left( \text{t} \right),\ldots, {{{\hat{x}}}_{N}}\left( \text{t} \right) \right]}^{T}}
\end{equation}
\begin{equation}
e\left( t \right)=\hat{x}\left( t \right)-x\left( t \right)
\nonumber\end{equation}
\begin{equation}
\delta \left( t \right)={{\left[ ~{{\delta }_{1}}\left( \text{t} \right),\ldots, {{\delta }_{N}}\left( \text{t} \right) \right]}^{T}}.
\nonumber
\label{eq_11}
\end{equation}
We can rewrite the closed-loop system from equations (\ref{eq_1}) and (\ref{eq_3}) in a compact form as 
\begin{equation}
    \begin{aligned}
x\left( t+1 \right)= ({{I}_{N-1}}\otimes A-{H}\otimes BK)\text{ }\!\!~\!\!\text{ }x\left( t \right)-\left( H\otimes BK \right)e\left( t \right)
    \end{aligned}
    \label{eq_12}
\end{equation}
Consequently, the consensus problem of a leader-follower discrete-time system has been converted to the stability problem of the error system.

Now, we provide the design procedure for the control law. Note that $\bar{\L}$ is the Laplacian matrix of the platoon with a leader for graph $\tilde{G}$  which are the eigenvalues of the matrix $\bar{\L}$ as $ 0={{\bar{\lambda }}_{1}}<{{\bar{\lambda }}_{2}}<\ldots <{\bar{\lambda }}_{N}<{{\bar{\lambda }}_{N+1}}$ .
Under Assumption 1, the parameter $\xi>0$  is chosen in such a way that
\begin{equation}
\underset{j}{\mathop \prod }\,\left| \lambda _{j}^{u}\left( A \right) \right|<{{\text{ }\!\!\xi\!\!\text{ }}^{-1}}<\frac{1+{{{{\bar{\lambda }}}_{2}}}/{{{{\bar{\lambda }}}_{\text{N+1}}}}}{1-{{{{\bar{\lambda }}}_{2}}}/{{{{\bar{\lambda }}}_{\text{N+1}}}}},
\label{eq_13}
\end{equation}
where $\left| \lambda _{j}^{u}\left( A \right) \right|$ denotes the unstable eigenvalues of ${A}$. 
Under Assumption 1, the following modified algebraic Riccati inequality in the discrete-time domain is solved to 
obtain a positive-definite matrix ${P}$, where
\begin{equation}
P-{{A}^{T}}PA+\left( 1-{{\text{ }\!\!\xi\!\!\text{ }}^{2}} \right)\frac{{{A}^{T}}PB{{B}^{T}}PA}{{{B}^{T}}PB}=\text{W}>0
\label{eq_14}
\end{equation}
and the following condition holds 
\begin{equation}
\frac{\left( 1-\text{ }\!\!\xi\!\!\text{ } \right)\left( {{{\bar{\lambda }}}_{2}}+{{{\bar{\lambda }}}_{N+1}} \right)}{2}<{{\lambda }_{i}}<\frac{\left( 1+\text{ }\!\!\xi\!\!\text{ } \right)\left( {{{\bar{\lambda }}}_{2}}+{{{\bar{\lambda }}}_{N+1}} \right)}{2}
\label{eq_15}
\end{equation}
for$~i=1,2,\ldots ,N$. Furthermore, The controller gain is designed as in [5], that is
\begin{equation}
K=\frac{2}{{{{\bar{\lambda }}}_{2}}+{{{\bar{\lambda }}}_{N+1}}}\frac{{{B}^{T}}PA}{{{B}^{T}}PB}
\label{eq_16}
\end{equation}

\subsection{Attack Model}
     Considering the influence of malicious attacks and the constraint of limited communication resources, defining affected and safe intervals in the presence of the attack and event-triggering sequences is crucial. In this section, a gain modification attack is introduced and formulated. In a gain modification attack, the attacker tries to alter the control gain of the vehicles by modifying the ${{k}_{v}}$ gain. The reason for choosing ${{k}_{v}}$ as the attack target is that its manipulation can cause more harm than that of ${{k}_{p}}$. The attacker can achieve this manipulation by constantly accelerating and braking. Also, we assume that attacks cannot affect the leader vehicle.

\emph{Lemma 1:}
The platoon system is Schur stable if and only if the below conditions are met,
\begin{equation}
0~<\hat{T}{{k}_{p}}<{{k}_{v}}<\frac{2}{\hat{T}{{\lambda }_{\text{max}}}}+\frac{\hat{T}{{k}_{p}}}{2}
\label{eq_17}
\end{equation}

\emph{Proof}: See Appendix A.

 Lemma 1 provides a crucial insight into the specific range of ${{k}_{v}}$ required for attackers to induce oscillation and destabilization from each position in the platoon. The determination of this minimum value takes into account the topology structure and dynamics of the vehicles in the platoon. 
The controller gain obtained from equation (\ref{eq_16}) is within the stability range defined by condition (\ref{eq_17}). According to equation (\ref{eq_17}), an attack on the platoon will cause instability if the controller gains ${{k}_{v}}$ exceeds $\frac{2}{\hat{T}{{\lambda }_{\text{max}}}}+\frac{\hat{T}{{k}_{p}}}{2}$ or is less than $\hat{T}{{k}_{p}}$ for all vehicles. However, it is still possible to have a stable platoon if one or some vehicles exceed this stability limit. To ensure safety, we consider the condition (\ref{eq_17}) as a conservative limit for any number of attacked vehicles.

The time sequence when an adversary launches a gain modification attack is denoted by ${{h}_{j}}$. The attack may last for a dwell time, denoted by ${\tau }_{j}$. During attack interval $[{{h}_{j}},{{h}_{j}}+{{\tau }_{j}})$, the system is under attack. The attacker tries to alter controller gains to push them outside the stability range (\ref{eq_17}) and render them inaccessible then the control protocols ${{u}_{i}(t)}$ of the agents remain unchanged during the attack.
Then, attackers may go into sleep mode and accumulate energy for the next launch. Define ${{\Psi }_{att}}\left( j \right)=[{{h}_{j}},{{h}_{j}}+{{\tau }_{j}}]$ as the $j$th attack duration time and let ${{\Psi }_{safe}}\left( j \right)=({{h}_{j}}+{{\tau}_{j}},{{h}_{j+1}}]$ be the consequent safe time. Here, $\Xi \left( {{t}_{0}},t \right)=\cup {{\Psi }_{att}}\cap \left[ {{t}_{0}},t \right]\text{ }\!\!~\!\!\text{ }$ denotes the union of the attack intervals during $\left[ {{t}_{0}},t \right)$. Furthermore, we assume that the attack duration and the attack frequency are constrained as follows.
 
\emph{Assumption 2 (Attack Duration)}: For ${\forall{t}}$, ${{t}_{0}}\ge 0$ with $t\ge {{t}_{0}}$, the attack duration time satisfies
\begin{equation}
\left| \Xi \left( {{t}_{0}},t \right) \right|\le {{\zeta }_{0}}+~{{\tau }_{0}}\left( ~t-{{t}_{0}} \right),
\label{eq_18}
\end{equation}
where $0<{{\tau }_{0}}<1$ and ${{\zeta }_{0}}\ge 0$.

\emph{Assumption 3 (Attack Frequency)}:
For ${\forall{t}}$, ${{t}_{0}}\ge 0$ with $t\ge {{t}_{0}}$. Let $F({{t}_{0}},t)$be the number of attacks launched during$\left[ {{t}_{0}},t \right)$. It satisfies
\begin{equation}
{F}\left( {{t}_{0}},t \right)\le {{\mathcal{F}}_{0}}+{{f}_{0}}\left( t-{{t}_{0}} \right),
\label{eq_19}
\end{equation}
where ${{f}_{0}}>0~$and ${{\mathcal{F}}_{0}}\ge 0$.

We should note that these assumptions are similar to those presented in \cite{de2015input,feng2019secure,he2021secure}.
 In this study, we have taken into consideration that the control input remains constant during the intervals between triggering times. As a result, we have observed that the impact on the controller's gain in the control input is only observed when a triggering event occurs. Next, we will focus on the relationship between the attack and triggering sequences. The time sequence $\left\{ {{T}_{k}} \right\}$ is a merged sequence that includes all the individual triggering sequences $t_{k}^{i}$ for each vehicle, arranged in chronological order, that is, $\left\{ {{T}_{k}} \right\}=\{t_{k}^{i}\,|\,i\in N,~k\in {{N}^{+}}\}$ and ${{T}_{0}}<{{T}_{1}}<\cdots <{{T}_{k}}<{{T}_{k+1}}<\cdots$ At least one vehicle is triggered at every instant ${{T}_{k}}$ and there is no event triggering during $\left[ {{T}_{k}},{{T}_{k+1}} \right)$.

 For vehicle$~i$, triggering sequence $\left\{ t_{k}^{i},t_{k+1}^{i},\ldots ,t_{k+m}^{i} \right\}$ represents the instances when the vehicle is under attack, it is evident that during the time interval $[t_{k}^{i},t_{k+m+1}^{i})$, a manipulated control input will be applied to the vehicle $i$. The dwell time of an attack, which represents the duration during which the control input is manipulated by the attacker, may not be the same as the affected dwell time of the vehicles. In the context of the system's affected duration induced by the $j$th attack, it is defined as ${{\tilde{\Psi }}_{att}}\left( j \right)=\left[ H_{j}^{att},H_{j}^{safe} \right]$ where $H_{j}^{att}$ represents the first triggering instant after the $j$th attack has been launched and $H_{j}^{safe}$ refers to the first instant after the attack has ceased. In the worst case, ${{\tilde{\Psi }}_{att}}\left( j \right)={{\Psi }_{att}}\left( j \right)+{{\Delta }^{\text{*}}}$, where ${{\Delta }^{\text{*}}}$ represents the maximum triggering interval affected by the attack. In this case, the attack affects the system at one triggering instant ${{T}_{k}}$ and ends at another triggering instant ${{T}_{l}}$. Thus, we define  $\tilde{\Xi }\left( {{k}_{0}},k \right)$ as the union of affected intervals belonging to $\left[ {{t}_{0}},t \right)$. Similar to \cite{he2021secure}, the relationship between the attack duration ${{\Psi }_{att}}\left( j \right)$ and the affected duration ${{\tilde{\Psi }}_{att}}\left( j \right)$ is provided. Let ${{\varrho }_{i}}\left( t \right)$ denote whether vehicle $i$ is attacked at time $t$. If ${{\varrho }_{i}}\left( t \right)=1$, the attack is launched for vehicle $i$, otherwise, ${{\varrho }_{i}}\left( t \right)=0$. The control input (\ref{eq_3}) under attack can be calculated as
 \begin{equation} 
       \begin{aligned}
 {{u}_{i}}\left( t_{k}^{i} \right)= \left(1-\varrho \left( t_{k}^{i} \right)\right)  K\left( t \right) {{\hat{q}}_{i}}\left(t_{k}^{i}\right)+ \varrho \left( t_{k}^{i} \right) K_{g}\left( t \right) {{\hat{q}}_{i}}\left(t_{k}^{i}\right) \\ ,~~~t\in \left[ t_{k}^{i}~,t_{k+1}^{i} \right)~ , ~~~~~ K_{g}\left( t \right)=\left( K\left( t \right)+{{{{g}}}_{v}}\left( t \right) \right)
       \label{eq_20}
           \end{aligned}
 \end{equation}	
Where $~~{{{{g}}}_{v}}=\left[ 0 , {{{\tilde{g}}}_{v}} \right] $ ,
$~{{{{g}}}_{v}}  \in {{\mathbb{R}}^{1\times 2}}$.
\\
\section{Event-triggered secure control}
In this section, we introduce two distributed protocols for the platoon system that offer different characteristics and advantages in terms of computational resources and communication frequency reduction. These protocols utilize an event-based triggering control scheme, where vehicles transmit their states only when the difference from the last transmitted data exceeds a predefined threshold. Both protocols determine the thresholds based on the locally available states of each agent. However, the difference between them lies in whether the thresholds incorporate dynamic auxiliary variables. Hence, they will be referred to as static and dynamic event-triggering protocols and are discussed in Sections A and B, respectively. 

\subsection{Secure static event-triggered consensus under sequential gain modification attack}
We develop an asynchronous static event-triggered control scheme to induce agents to achieve a common state. The triggering function is described by
\begin{equation}
\label{eq_21}
\left\Vert{{e}_{l}}\right\Vert^2-\frac{\partial \left( {{s}_{1}}-{{s}_{2}}\beta  \right)}{({{s}_{2}}+{{s}_{3}}\beta ){{\beta }^{-1}}} \left\Vert{\hat{q}_{l}}\right\Vert^2<0
\end{equation}
Where $0<\partial<1$. Also, there exists a positive number $\beta $ that satisfies the inequality $\beta <\frac{{{s}_{1}}}{{{s}_{2}}}$ where positive numbers ${{s}_{1}},{{s}_{2}},{{s}_{3}}$ are denoted as follows:
\begin{equation}
{{s}_{1}}=\min \left[ {{W}_{1}}~{{\tau }_{min}}\left( {H}^{-2}\otimes {{I}_{N}} \right) \right]
\label{eq_22}
\end{equation}
\begin{equation}
    \begin{aligned}
{{s}_{2}}=\max \left\{ \left[ {{\tau }_{max}}{{\left( {{I}_{N}}\otimes Bk \right)}^{T}}\left( {{I}_{N}}\otimes P \right) \right. \right. \\ 
 \break \times \left. \left( {{I}_{N}}\otimes A-{H}\otimes BK \right) \right] \\ 
 \break \left. -\left( {H}^{-1}\otimes {{W}_{1}}{{I}_{N}} \right) \right\}
    \end{aligned}
    \label{eq_23}
\end{equation}
\begin{equation}
    \begin{aligned}
{{s}_{3}}=\max \left\{ \left[ {{\tau }_{max}}{{\left( {H}\otimes Bk \right)}^{T}}\left( {{I}_{N}}\otimes P \right) \right. \right. \\ 
 \break \times \left. \left( {{I}_{N}}\otimes 2A-{H}\otimes BK \right) \right]-{{W}_{1}} \left. \right\}
    \end{aligned}
    \label{eq_24}
\end{equation}
Where $W={{W}_{1}}+{{W}_{2}}$ and $ 0 < {{W}_{1}}< W$ , $\text{ }\!\!~\!\!\text{ }{{W}_{1}},{{W}_{2}}>0$ which implies that ${{s}_{2}> 0}$, ${{s}_{3}> 0}$.

\emph{Theorem 1}: Consider the homogeneous platoon system (\ref{eq_1}) with the leader (\ref{eq_2}) satisfies Assumptions 1–3. Under the static event-triggered controller (\ref{eq_3}) and the gain matrix (\ref{eq_16}), for any initial condition, subject to a sequential gain modification attack, the consensus of the platoon can be asymptotically achieved with the triggering function (\ref{eq_21}) if the following condition holds.
\begin{equation}
{{\tau }_{0}}+{{\Delta }^{*}}{{f}_{0}}<\frac{{\tilde{\alpha }}}{\tilde{\alpha }+\tilde{\gamma}}
\label{eq_25}
\end{equation}	
Where ${{\Delta }^{*}}$ is the maximum affected triggering interval and

\begin{equation}
\label{eq_26}
\tilde{\alpha }=\frac{[\left( 1-\partial\right)({{s}_{1}}-\beta {{s}_{2}}){{\left({{\lambda }_{N}}^{-1}-\sqrt{\frac{\partial \left({{s}_{1}}-{{s}_{2}}\beta\right)}{({{s}_{2}}+{{s}_{3}}\beta ){{\beta }^{-1}}}} \right)}^{-2}}
+{{\text{W}}_{2}}]}{{{\lambda}_{N}}\left(P\right)}
\end{equation}
\begin{equation}
\label{eq_27}
\tilde{\beta}=\frac{-[\left( 1-\partial  \right)~({{s}_{1}}-\beta {{\tilde{s}}_{2}}){{\left( {{\lambda }_{N}}^{-1}-\sqrt{\frac{\partial \left( {{s}_{1}}-{{{\tilde{s}}}_{2}}\beta  \right)}{({{{\tilde{s}}}_{2}}+{{{\tilde{s}}}_{3}}\beta ){{\beta }^{-1}}}} \right)}^{-2}}+{{\text{W}}_{2}}]}{{{\lambda}_{N}}\left(P\right)}
\end{equation}

\emph{Proof}: Consider the following Lyapunov function as

\begin{equation}
V\left( t \right)={{\delta }^{T}}\left( t \right)\left( {{I}_{N}}\otimes \text{P} \right)\delta ~\left( t \right)
\label{eq_28}
\end{equation}

Using (\ref{eq_28}) and taking the difference between the Lyapunov function at times $t+1$ and $t$ we will have:

\[
\begin{aligned}
\nabla V &= V(t+1) - V(t) \\
\end{aligned}
\]
\[
\begin{aligned}
&= \delta^T(t+1) (I_N \otimes P) \delta(t+1) - \delta^T(t) (I_N \otimes P) \delta(t)
\end{aligned}
\]
\[
\begin{aligned}
&= (\delta^T(t) (I_N \otimes A - H \otimes B K)^T - e^T(H \otimes B K)^T) \notag \\
\end{aligned}
\]
\[
\begin{aligned}
&\quad \times (I_N \otimes P) ((I_N \otimes A - H \otimes B K) \delta(t) - (H \otimes B K) e) \notag \\
\end{aligned}
\]
\[
\begin{aligned}
&\quad - \delta^T(t) (I_N \otimes P) \delta(t) \notag \\
\end{aligned}
\]
\[
\begin{aligned}
&= \delta^T(t) ((I_N \otimes A - H \otimes B K)^T (I_N \otimes P) \notag \\
\end{aligned}
\]
\[
\begin{aligned}
&\quad \times (I_N \otimes A - H \otimes B K) - (I_N \otimes P)) \delta(t) \notag \\
\end{aligned}
\]
\[
\begin{aligned}
&\quad - \delta^T(t) (I_N \otimes A - H \otimes B K)^T (I_N \otimes P) (H \otimes B K) e \notag \\
\end{aligned}
\]
\[
\begin{aligned}
&\quad - e^T (H \otimes B K)^T (I_N \otimes P) (I_N \otimes A - H \otimes B K) \delta(t) \notag \\
\end{aligned}
\]
\[
\begin{aligned}
&\quad + e^T (H \otimes B K)^T (I_N \otimes P) (H \otimes B K) e \notag \\
\end{aligned}
\]
\[
\begin{aligned}
&\le -W \delta^T(t) \delta(t) - 2 e^T (H \otimes B K)^T (I_N \otimes P) \notag \\
\end{aligned}
\]
\[
\begin{aligned}
&\quad \times (I_N \otimes A - H \otimes B K) \delta(t) + e^T (H \otimes B K)^T (I_N \otimes P) \notag \\
\end{aligned}
\]
\[
\begin{aligned}
&\le - (W_2 + W_1) \{ \hat{q}^T (H^{-2} \otimes I_N) \hat{q} + 2 \hat{q}^T (H^{-1} \otimes I_N) e + e^T e \} \notag \\
\end{aligned}
\]
\[
\begin{aligned}
&\quad - 2 e^T (H \otimes B K)^T (I_N \otimes P) \notag \\
\end{aligned}
\]
\[
\begin{aligned}
&\quad \times ((I_N \otimes A - H \otimes B K) (- (H^{-1} \otimes I_N) \hat{q} - e)) \notag \\
\end{aligned}
\]
\begin{equation}
 + e^T (H \otimes B K)^T (I_N \otimes P) (H \otimes B K)e 
\label{eq_29}
\end{equation}

Thus

\[
\begin{aligned}
&\quad\nabla V=V\left( t+1 \right)-V\left( t \right) \notag \\
\end{aligned}
\]
\[
\begin{aligned}
&\quad =-{{\text{W}}_{2}}{{\delta }^{T}}\left( t \right)\delta \left( t \right)-{{\text{W}}_{1}}{{\hat{q}}^{T}}\left( {H}^{-2}\otimes {{I}_{N}} \right)\hat{q} \notag \\
\end{aligned}
\]
\[
\begin{aligned}
&\quad -{{\text{W}}_{1}}{{e}^{T}}e 
-2{{\text{W}}_{1}}{{\hat{q}}^{T}}\left( {H}^{-1}\otimes {{I}_{N}} \right)e \notag \\
\end{aligned}
\]
\[
\begin{aligned}
&\quad +2{{e}^{T}}{{\left( {H}\otimes BK  \right)}^{T}}\left( {{I}_{N}}\otimes \text{P} \right)
\left( {{I}_{N}}\otimes \text{A}-{H}\otimes B K \right)e
& \notag 
\end{aligned}
\]
\begin{equation}
 +{{e}^{T}}{{\left( {H}\otimes B K  \right)}^{T}} \left( {{I}_{N}}\otimes \text{P} \right)\left( {H}\otimes B K \right)e 
\label{eq_30}
\end{equation}

Where ${{\text{W}}_{2}}>0$ satisfies $\text{W}={{\text{W}}_{1}}+{{\text{W}}_{2}}$. We know that
$~~{{m}^{T}}n\le \left\Vert{m}\right\Vert\left\Vert{n}\right\Vert$ then we can achieve
\[
\begin{aligned}
&\quad-{{\text{W}}_{1}}{{\hat{q}}^{T}}\left( {{L}_{g}}^{-2}\otimes {{I}_{N}} \right)\hat{q}
\le -{{s}_{1}}\left\Vert{\hat{q}}\right \Vert^{2}, \notag 
\end{aligned}
\]
\[
\begin{aligned}
&\quad -2{{\text{W}}_{1}}{{\hat{q}}^{T}}\left( {H}^{-1}\otimes {{I}_{N}} \right)e+2{{e}^{T}}{{\left( {{I}_{N}}\otimes BK \right)}^{T}}\left( {{I}_{N}}\otimes \text{P} \right)
\left( {{I}_{N}}\otimes \text{A}-{H}\otimes BK \right)\hat{q}
\le +2{{s}_{2}}\left\Vert\hat{q}\right\Vert\left\Vert{e}\right\Vert, \notag \\
\end{aligned}
\]
\[
\begin{aligned}
&\quad -{{\text{W}}_{1}}{{e}^{T}}e+2{{e}^{T}}{{\left( {H}\otimes B K  \right)}^{T}}
\left( {{I}_{N}}\otimes \text{P} \right)
\left( {{I}_{N}}\otimes \text{A}-{H}\otimes BK \right)e
+{{e}^{T}}{{\left( {H}\otimes BK \right)}^{T}}\left( {{I}_{N}}\otimes \text{P} \right)
\left( {H}\otimes B K  \right)e \notag 
\end{aligned}
\]
\begin{equation}
\le +{{s}_{3}}{\left\Vert{e}\right\Vert^2} \notag
\label{eq_32}
\end{equation}

The equation (\ref{eq_30}) can be rewritten as follows:
\begin{equation} 
\begin{split}
&V\left( t+1 \right)-V\left( t \right)
\le -{{s}_{1}}\left\Vert\hat{q}\right\Vert^2~+2{{s}_{2}}\left\Vert\hat{q}\right\Vert\left\Vert{e}\right\Vert \\
&\break+{{s}_{3}}\left\Vert{e}\right\Vert^2-{{\text{W}}_{2}}{{\delta }^{T}}\left( t \right)\delta \left( t \right)
\end{split}
\label{eq_31}
\end{equation} 
Using Young’s inequality ${{m}^{T}}n\le \frac{1}{2\beta }\left\Vert{m}\right\Vert^2+\frac{\beta }{2}\left\Vert{n}\right\Vert^2 $, for any $\beta >0$, we obtain  

\[
\begin{aligned}
&\quad V\left( t+1 \right)-V\left( t \right) \le \notag \\
\end{aligned}
\]
\[
\begin{aligned}
&\quad -{{s}_{1}}\left\Vert{q}\right\Vert^2+{{s}_{2}}~\beta \left\Vert\hat{q}\right\Vert^2\\
&+{{s}_{2}}\frac{1}{\beta }\left\Vert{e}\right\Vert^2
+{{s}_{3}}\left\Vert{e}\right\Vert^2-{{\text{W}}_{2}}{{\delta }^{T}}\left( t \right)\delta \left( t \right) \notag \\
\end{aligned}
\]
\begin{equation}
 \le -[({{s}_{1}}-\beta {{s}_{2}})\left\Vert\hat{q}\right\Vert^2-\left( \frac{{{s}_{2}}}{\beta }
+{{s}_{3}} \right)\left\Vert{e}\right\Vert^2]-{{\text{W}}_{2}}
)\left\Vert{\delta}\right\Vert^2  
\label{eq_32}
\end{equation}

According to the static event-triggering condition (\ref{eq_21}), one has:
\begin{equation} 
\begin{split}
&V\left( t+1 \right)-V\left( t \right)\\
&\le -[\left( 1-\partial  \right)~({{s}_{1}}-\beta {{s}_{2}})~\left\Vert\hat{q}\right\Vert^2~]-{{\text{W}}_{2}}{{\delta }^{T}}\delta
\end{split}
\label{eq_33}
\end{equation}

The above equation holds coming from the fact that as $H$ is symmetric and positive, there exists an orthogonal matrix U such that$~\delta =\left( U\otimes {{I}_{N}} \right)\hat{\delta }$ with ${{U}^{T}}{U}={{I}_{N}}$. It is easy to know that ${{U}^{T}}{H}U=diag\left\{ {{\lambda }_{1}}\left( {H} \right),~\ldots ~,~{{\lambda }_{N}}\left( {H} \right) \right\}$. Let $\hat{q}\left( t \right)\le col\left( {{{\hat{q}}}_{1}},\ldots ,~{{{\hat{q}}}_{N}} \right)$ with ${{\hat{q}}_{l}}=\underset{j\in {{N}_{l}}}{\mathop \sum }\,\left( \left( {{{\hat{x}}}_{j}}-{{{\hat{x}}}_{l}} \right)+m\left( {{x}_{0}}-{{{\hat{x}}}_{l}} \right) \right)$ and ${{e}_{l}}\left( t \right)=~{{\hat{x}}_{l}}\left( t \right)-~{{x}_{l}}\left( t \right)$
we have
\[
\begin{aligned}
&\quad  \left\Vert\hat{q}\right\Vert=-\left\Vert\left( {H}\otimes {{I}_{N}} \right)\left( \hat{x}-{{x}_{0}} \right)\right\Vert  \notag \\
\end{aligned}
\]
\[
\begin{aligned}
&\quad =-\left\Vert\left( {H}\otimes {{I}_{N}} \right)\left( x-{{x}_{0}}+e \right)\right\Vert =\left\Vert{q}-\left( {H}\otimes {{I}_{N}} \right){e}\right\Vert  \notag \\
\end{aligned}
\]
\[
\begin{aligned}
&\quad   \le \left\Vert{q}\right\Vert+\left\Vert\left( {H}\otimes {{I}_{N}} \right){e}\right\Vert  \notag \\
\end{aligned}
\]
\begin{equation}
\le \left\Vert{q}\right\Vert+{{\lambda }_{N}}\left( {H} \right)\left\Vert{e}\right\Vert 
\label{eq_37}
\end{equation}

Then we know
\begin{equation} 
\begin{split}
\left\Vert{q}\right\Vert^2={{\left( x-{{x}_{0}} \right)}^{T}}{{\left( {H}\otimes {{I}_{N}} \right)}^{T}}\left( {H}\otimes {{I}_{N}} \right)\left( x-{{x}_{0}} \right)\\
={{\delta }^{T}}{{\left( {H}\otimes {{I}_{N}} \right)}^{2}}\delta \le {{\lambda }_{N}}^{2}{{\left( {H} \right)}}~\left\Vert{\delta}\right\Vert^2
\end{split}
\label{eq_38}
\end{equation}
Based on equations (\ref{eq_37}) and (\ref{eq_38}), we can get
$\left\Vert\hat{q}\right\Vert\le {{\lambda }_{N}}\left( {H} \right)~ \left( \left\Vert{\delta}\right\Vert +\left\Vert{e}\right\Vert \right)$, which leads to
	$\left\Vert{e}\right\Vert>{{\lambda }_{N}}^{-1}\left\Vert\hat{q}\right\Vert-\left\Vert{\delta}\right\Vert$, by combining with (\ref{eq_21}) yields
 
\begin{equation} 
\begin{split}
&{{\lambda }_{N}}^{-1}\left\Vert\hat{q}\right\Vert-\left\Vert{\delta}\right\Vert <\sqrt{\frac{\partial \left( {{s}_{1}}-{{s}_{2}}\beta  \right)}{({{s}_{2}}+{{s}_{3}}\beta ){{\beta }^{-1}}}}\left\Vert\hat{q}\right\Vert~\\
 &\left\Vert\hat{q}\right\Vert<{{\left( {{\lambda }_{N}}^{-1}-\sqrt{\frac{\partial \left( {{s}_{1}}-{{s}_{2}}\beta  \right)}{({{s}_{2}}+{{s}_{3}}\beta ){{\beta }^{-1}}}} \right)}^{-1}}\left\Vert{\delta}\right\Vert
\end{split}
\label{eq_40}
\end{equation}
Now, by inserting equation (\ref{eq_40}) into the Lyapunov function (\ref{eq_33}), we obtained that
\begin{equation} 
\begin{split}
&V\left( t+1 \right)-V\left( t \right)\le -[\left( 1-\partial  \right)~({{s}_{1}}-\beta {{s}_{2}})\\
&\times{{\left( {{\lambda }_{N}}^{-1}-\sqrt{\frac{\partial \left( {{s}_{1}}-{{s}_{2}}\beta  \right)}{({{s}_{2}}+{{s}_{3}}\beta ){{\beta }^{-1}}}} \right)}^{-2}}+{{\text{W}}_{2}}]\left\Vert{\delta}\right\Vert^2
\end{split}
\label{eq_41}
\end{equation}

First, we consider the case that the system is free of attack, that is, $t\in \left[ {{T}_{k}},{{T}_{k+1}} \right],~{{T}_{k}}\notin \Xi \left( {{t}_{0}}~,~t \right)$. In such a case ${{\varrho }_{i}}\left( k \right)=0$
From (\ref{eq_41}), one has
\[
\begin{aligned}
&\quad V\left( t+1 \right)-V\left( t \right)
\le  \notag \\
\end{aligned}
\]
\[
\begin{aligned}
&\quad -[\left( 1-\partial  \right)~({{s}_{1}}-\beta {{s}_{2}})\\
&\times{{\left( {{\lambda }_{N}}^{-1}-\sqrt{\frac{\partial \left( {{s}_{1}}-{{s}_{2}}\beta  \right)}{({{s}_{2}}+{{s}_{3}}\beta ){{\beta }^{-1}}}} \right)}^{-2}} +{{\text{W}}_{2}}]~\left\Vert{\delta}\right\Vert^2   \notag \\
\end{aligned}
\]
\begin{equation}
 \le -\tilde{\alpha }~{{\delta }^{T}}\left( t \right)\left( {{I}_{N}}\otimes \text{P} \right)\delta ~\left( t \right) \le -\tilde{\alpha }~V\left( t \right) 
\label{eq_42}
\end{equation}

According to this theorem, the following conditions are met. 
\begin{equation}
({{s}_{1}}-\beta {{s}_{2}})>0
\label{eq_43}
\end{equation}
\begin{equation}
\left\Vert{{e}_{l}}\right\Vert^2<\frac{\partial(t) \left( {{s}_{1}}-{{s}_{2}}\beta  \right)}{({{s}_{2}}+{{s}_{3}}\beta ){{\beta }^{-1}}}\left\Vert{\hat{q}_{l}}\right\Vert^2
\label{eq_44}
\end{equation}	
Then, One can observe that $V\left( t+1 \right)-V\left( t \right)<0$, that lead to:
$\tilde{\alpha }>0$ 
Therefore
\begin{equation} 
\begin{split}
V\left( t \right)<{{e}^{-~\tilde{\alpha }\left( t-{{T}_{k}} \right)}}V\left( {{T}_{k}} \right)
\end{split}
\label{eq_45}
\end{equation}
for $~t\in \left[ {{T}_{k}},{{T}_{k+1}} \right],~{{T}_{k}}\notin \Xi \left( {{t}_{0}}~,~t \right)$ Due to $\tilde{\alpha }$ being a positive number. Thus, it follows that $V\left( t \right)\overset{~}{\mathop{\to }}\,0$ as t$\overset{~}{\mathop{\to }}\,\infty $. This implies that platoon system consensus can be asymptotically achieved. Hence, $~\left\Vert{\delta}\right\Vert^2=0$  as$~t\overset{~}{\mathop{\to }}\,\infty $.
\\Second, we are going to analyze the stability as$\text{ }\!\!~\!\!\text{ }t\in \left[ {{T}_{{{k}'}}},{{T}_{{k}'+1}} \right],\text{ }\!\!~\!\!\text{ }{{T}_{{{k}'}}}\in \Xi \left( {{t}_{0}}\text{ }\!\!~\!\!\text{ },\text{ }\!\!~\!\!\text{ }t \right)$, which is subject to sequential gain modification attacks. When$~{{\varrho }_{i}}\left( t \right)=1$, a attack is launched. The change in the gain controller can cause variations in the parameters of $s_2$ and $s_3$. Consequently, this may result in equation (\ref{eq_43}) not being satisfied. 

One can derive from \ref{eq_41} and the above parameters that
\begin{equation}
    \begin{aligned}
V\left( t+1 \right)-V\left( t \right)\le -\left[ \left( 1-\partial  \right){{\left( s\right.}_{1}}-\beta {{{\tilde{s}}}_{2}} \right)\\
\break{\left( {{\lambda }_{N}}^{-1}-\sqrt{\frac{\partial \left( {{s}_{1}}-{{{\tilde{s}}}_{2}}\beta  \right)} {\left. {{\left( {\tilde{s}} \right.}_{2}}+{{{\tilde{s}}}_{3}}\beta  \right){{\beta }^{-1}}}} \right)}^{-2}+{{\tilde{W}}_{2}} 
]\left\Vert{\delta}\right\Vert^2\le\\
\break \tilde{\gamma} {{\delta }^{T}}\left( k \right)\left( {{I}_{N}}\otimes \text{P} \right)\delta ~\left( k \right)\le \tilde{\gamma} V\left( t \right)
    \end{aligned}
    \label{eq_48}
\end{equation}
Thus
\begin{equation}
    \begin{aligned}
V\left( t \right)<{{e}^{-\tilde{\gamma}\left( t-{{T}_{{{k}'}}} \right)}}V\left( {{T}_{{{k}'}}} \right),
    \end{aligned}
    \label{eq_49}
\end{equation}
for $t\in \left[ {{T}_{{{k}'}}},{{T}_{{k}'+1}} \right]~~,~~{{T}_{{{k}'}}}\in {\Xi }\left( {{t}_{0}}~,~t \right)$. Now we are in the position to consider V(t) for$\text{ }\!\!~\!\!\text{ }\left( {{t}_{0}}~,~t \right)$. It is obvious that \ref{eq_45} is activated when $t\notin \tilde\Xi \left( {{t}_{0}}~,~t \right)$, while \ref{eq_49} is activated when $t\in \tilde\Xi \left( {{t}_{0}}~,~t \right)$. For $t\in \left[ H_{j-1}^{safe},H_{j}^{att} \right]$, we have
\begin{equation}
    \begin{aligned}
V\left( k \right)\le {{e}^{-\tilde{\alpha }\left( t-H_{j-1}^{safe} \right)}} V\left( H_{j-1}^{safe} \right)\le {{e}^{-\tilde{\alpha }\left( t-_{j-1}^{safe} \right)}}\\
\break{{e}^{\tilde{\gamma}\left( H_{j-1}^{safe}-H_{j}^{att} \right)}}V\left( H_{j-1}^{att} \right)\le \ldots
\\
\le {{e}^{-\tilde{\alpha }\left( t-{{t}_{0}}-\left| \tilde{\Xi }\left( {{t}_{0}},t \right) \right| \right)}}{{e}^{\tilde{\gamma}\left( \left| \tilde{\Xi }\left( {{t}_{0}},t \right) \right| \right)}}V\left( {{t}_{0}} \right)
    \end{aligned}
    \label{eq_51}
\end{equation}
Similarly, for$~t\in \left[ H_{j}^{att},H_{j-1}^{safe} \right]$ we can get
\begin{equation}
    \begin{aligned}
V\left( k \right)\le {{e}^{-\tilde{\gamma}\left( t-H_{j}^{att} \right)}}V\left( H_{j}^{att} \right) \\
\le {{e}^{\tilde{\gamma}\left( t-H_{j-1}^{att} \right)}}{{e}^{-\tilde{\alpha }\left( H_{j}^{att}-H_{j-1}^{safe} \right)}}V\left( H_{j-1}^{safe} \right) \\
\le {{e}^{-\tilde{\alpha }\left( t-{{t}_{0}}-\left| \tilde{\Xi }\left( {{t}_{0}},t \right) \right| \right)}}{{e}^{\tilde{\gamma}\left( \left| \tilde{\Xi }\left({{t}_{0}},t\right) \right| \right)}}V\left( {{t}_{0}} \right)~~~~ 
    \end{aligned}
    \label{eq_52}
\end{equation}
Thus, for $\forall t>{{t}_{0}}$, it follows from \ref{eq_51} and \ref{eq_52} that
\begin{equation}
    \begin{aligned}
V\left( k \right)\le {{e}^{-\tilde{\alpha }\left( t-{{t}_{0}}-\left| \tilde{\Xi }\left( {{t}_{0}},t \right) \right| \right)}}{{e}^{\tilde{\gamma}\left( \left| \tilde{\Xi }\left( {{t}_{0}},t \right) \right| \right)}}V\left( {{t}_{0}} \right)
    \end{aligned}
    \label{eq_53}
\end{equation}
Note that based on ${{\tilde{\Psi }}_{att}}\left( j \right)\le {{\Psi }_{att}}+{{\Delta}^{*}}$, one obtains 
\begin{equation}
    \begin{aligned}
\tilde{\Xi }\left( {{t}_{0}},t \right)\le \left| \Xi \left( {{t}_{0}},t \right) \right|+{{\Delta}^{*}}  \text{F}\left( {{t}_{0}},t \right)
    \end{aligned}
    \label{eq_54}
\end{equation}

According to Assumptions 2 and 3 and \ref{eq_54}, it follows from \ref{eq_53} that:
\begin{equation}
    \begin{aligned}
V\left( t \right)\le V\left( {{t}_{0}} \right){{e}^{-\tilde{\alpha }\left( t-{{t}_{0}}-\left| \tilde{\Xi }\left( {{t}_{0}},t \right) \right| \right)+\tilde{\gamma}\left( \left| \tilde{\Xi }\left( {{t}_{0}},t \right) \right| \right)}} \\
\break \le V\left( {{t}_{0}} \right){{e}^{-\tilde{\alpha }\left( t-{{t}_{0}} \right)}}{{e}^{\left( \tilde{\alpha }+\tilde{\gamma} \right)\left( \left| \tilde{\Xi }\left( {{t}_{0}},t \right) \right| \right)}} \\
\break \le V\left( {{t}_{0}} \right){{e}^{-\tilde{\alpha }\left( t-{{t}_{0}} \right)}}{{e}^{\left( \tilde{\alpha }+\tilde{\gamma} \right)\left( {{\zeta }_{0}}+~{{\tau }_{0}}\left( ~t-{{t}_{0}} \right)+{{\Delta}^{*}}\left( {{\mathcal{F}}_{0}}+{{f}_{0}}\left( ~t-t \right) \right) \right)}} \\
\break \le V\left( {{t}_{0}} \right){{e}^{-\tilde{\alpha }\left( t-{{t}_{0}} \right)}}{{e}^{\left( \tilde{\alpha }+\tilde{\gamma} \right)\left( {{\zeta }_{0}}+{{\Delta}^{*}}{{\mathcal{F}}_{0}}+\left( ~{{\tau }_{0}}+{{\Delta}^{*}}{{f}_{0}} \right)\left( ~t-{{t}_{0}} \right) \right)}} \\
\break \le V\left( {{t}_{0}} \right)~{{e}^{\left( \tilde{\alpha }+\tilde{\gamma} \right)\left( {{\zeta }_{0}}+{{\Delta}^{*}}{{\mathcal{F}}_{0}} \right)}}~{{e}^{\left[ -\tilde{\alpha }+~\left( \tilde{\alpha }+\tilde{\gamma} \right)({{\tau }_{0}}+{{\Delta}^{*}}{{f}_{0}}) \right]\left( ~t-{{t}_{0}} \right)}}
    \end{aligned}
    \label{eq_55}
\end{equation}

By (\ref{eq_25}), let$\text{ }\!\!~\!\!\text{ }-\tilde{\alpha }+( \tilde{\alpha }+\tilde{\gamma} )({{\tau }_{0}}+{{\Delta }^{*}}{{f}_{0}})<0$, so the inequality in (\ref{eq_55}) implies that $V(t)$ is bounded and error system is asymptotically stable. The proof is finished.

\emph{Remark 1}: According to Theorem 1, it has been proven that platoon systems can achieve secure consensus exponentially under the constraint of condition (\ref{eq_25}) on the attack duration and frequency within certain bounds.

\emph{Remark 2}: All controller gains are without attack when an attack is inactive. In this case, equation (\ref{eq_25}) is established for stability, indicating that the system can be regarded as being in a stable mode with an exponential convergence rate denoted by $\tilde{\alpha}$. However, when the attack is activated, the system's performance is influenced by the modification of the controller gain. This modification can lead to an increase in ${s}_{2}$, causing equation (\ref{eq_43}) to no longer be satisfied. As a result, the divergence rate $\tilde{\gamma}$ can be either positive or negative within the affected interval that the error system can exhibit stability or instability depending on the values of the modified gain.

\emph{Remark 3}: In Theorem 1, it is observed that there exists a trade-off between the convergence rate and the constraints imposed by the attacks. From equations (\ref{eq_18}), (\ref{eq_19}), and (\ref{eq_25}), we can conclude that if $\tilde{\alpha}$ increases or $\tilde{\gamma}$ decreases, the upper bounds on the attack duration and frequency will increase. This means that the system becomes more resilient. Conversely, if $\tilde{\alpha}$ decreases or $\tilde{\gamma}$ increases, the upper bounds on the attack duration and frequency will decrease. Overall, this trade-off implies that the platoon system exhibits resilience in achieving asymptotic consensus against sequential attacks. 

\subsection{Dynamic event-triggered secure consensus under sequential gain modification attacks}
In this section, we extend the secure static event-triggered protocol from the previous section to a new controller with a dynamic event-triggered protocol to deal with gain modification attacks. The vehicles determine their triggering instants $t_{k}^{i}~$by the following dynamic triggering function
\begin{equation}  
    \begin{aligned}
{{\theta }_{i}}\left( ~\left\Vert{{e}_{l}}\right\Vert^2-\frac{\partial \left( {{s}_{1}}-{{s}_{2}}\,\beta  \right)}{\left. {{\left( s \right.}_{2}}+{{s}_{3}}\beta\right){{\beta}^{-1}}} \left\Vert{\hat{q}_{l}}\right\Vert^2\right)
> {{\mu }_{i}}\left( t \right),~\forall \text{t}\in t_{k}^{i}
    \end{aligned}
    \label{eq_56}
\end{equation}

Where $\theta $ is a positive constant, and ${{\mu }_{i}}\left( t \right)$ is an internal dynamic variable satisfying the update rule given by
\begin{equation}
    \begin{aligned}
&{{\mu }_{i}}\left( t+1 \right)=\left(1-{{\rho }_{i}}\right){{\mu }_{i}}\left( t \right)\\
&+{{\vartheta }_{i}}\left( \frac{\partial \left( {{s}_{1}}-{{s}_{2}}\beta  \right)}{({{s}_{2}}+{{s}_{3}}\beta ){{\beta }^{-1}}}\left\Vert{\hat{q}_{l}}\right\Vert^2-\left\Vert{{e}_{l}}\right\Vert^2 \right)
    \end{aligned}
    \label{eq_57}
\end{equation}
where initial value is set as ${{\mu }_{i}}\left( {{t}_{0}} \right)>0$ and the parameters $~0<~{{\rho }_{i}}<1$ and $~~0<{{\vartheta }_{i}}<1$ are to be designed. The design parameters ${{\rho }_{i}},\vartheta $ and ${{\theta }_{i}}$ are chosen to satisfy the following conditions \cite{mishra2021average}
\begin{equation}
    \begin{aligned}
{{\theta }_{i}}>\frac{1-\vartheta }{{{\rho }_{i}}}   ~~~    ,   ~~~~        \vartheta +{{\rho }_{i}}<1     ~~~   , ~~     \vartheta <{{\theta }_{i}}    {{\rho }_{i}}
    \end{aligned}
    \label{eq_58}
\end{equation}

From equations (\ref{eq_56}) and (\ref{eq_57}), we have
\begin{equation}
    \begin{aligned}
{{\mu }_{i}}\left( t+1 \right)>\left( 1-{{\rho }_{i}}-\frac{{{\theta }_{i}}}{\text{ }\!\!~\!\!\text{ }\vartheta }~ \right){{\mu }_{i}}\left(t \right)
    \end{aligned}
    \label{eq_59}
\end{equation}
Hence, under the conditions in (\ref{eq_58}), we can show that ${{\mu }_{i}}\left( t \right)$ is always positive and
\begin{equation}
    \begin{aligned}
{{\mu }_{i}}\left( t \right)\ge {{\left( 1-{{\rho }_{i}}-\frac{{{\theta }_{i}}}{\text{ }\!\!~\!\!\text{ }\vartheta }~ \right)}^{t}}{{\mu }_{i}}\left( 0 \right)>0
    \end{aligned}
    \label{eq_60}
\end{equation}

It is clear that ${{\mu }_{i}}\left( 0 \right)>0$. as a result, ${{\mu }_{i}}\left( t \right)>0$ holds for [0,$\infty$).
This conclusion helps us to mitigate the number of triggering times compared with the static-triggering control.

\emph{Theorem 2}: Consider the homogeneous platoon system (\ref{eq_1}) with the leader (\ref{eq_2}) satisfies Assumptions 1–3. Under the dynamic event-triggered controller (\ref{eq_3}) and the gain matrix (\ref{eq_16}), for any initial condition subject to a gain modification attack, the consensus of the platoon can be asymptotically
achieved with the triggering function (\ref{eq_56}), and if there exists:

\begin{equation}
\frac{{{s}_{2}}}{\beta }+{{s}_{3}}>\vartheta
\label{eq_61}
\end{equation}
\begin{equation}
    \begin{aligned}
-{{(\theta }_{i}}\ {{\rho }_{i}} )+\vartheta >{{{s}}_{2}}{{\beta }^{-1}}+{{{s}}_{3}} 
    \end{aligned}
    \label{eq_62}
\end{equation}
\begin{equation}
    \begin{aligned}
        \tilde{\Gamma}=min\left\{ {\tilde{\alpha}} ,{{\alpha }_{1}} \right\}>0
    \end{aligned}
    \label{eq_63}
\end{equation}
\begin{equation}
    \begin{aligned}
  {{\tau }_{0}}+{{\Delta }^{*}}{{f}_{0}}<\frac{\tilde{\Gamma}}{\tilde{\Gamma}+\tilde{\gamma}}
    \end{aligned}
    \label{eq_64}
\end{equation}

Where ${{\Delta}^{*}}$ is the maximum affected triggering interval and parameters ${{s}_{l}}$, l = 1, 2, 3 are similar to the ones in theorem 1.

\emph{Proof}: Choose the following Lyapunov function:
\begin{equation}
    \begin{aligned}
{{V}_{1}}\left( t \right)={{\delta }^{T}}\left( t \right)\left( {{I}_{N}}\otimes \text{P} \right)\delta ~\left( t \right)\\
{{V}_{2}}\left( t \right)=\sum {{\mu }_{i}}\left( t \right)~~~~~~~~~~~~~~
\end{aligned}
\label{eq_65}
\end{equation}            
Calculating the time derivative of ${{V}_{2}}\left( t \right)$ yields
\begin{equation}
    \begin{aligned}
\nabla {{V}_{2}}={{\mu }_{i}}\left( t+1 \right)-{{\mu }_{i}}\left( t \right)=\left( 1-{{\rho }_{i}} \right){{\mu }_{i}}\left( t \right)\\
+\vartheta \left( \frac{\partial \left( {{s}_{1}}-{{s}_{2}}\beta  \right)}{({{s}_{2}}+{{s}_{3}}\beta ){{\beta }^{-1}}}\left\Vert{{q}_{l}}\right\Vert^2-\left\Vert{{e}_{l}}\right\Vert^2 \right)-~{{\mu }_{i}}\left( t \right)
    \end{aligned}
    \label{eq_67}
\end{equation}
Where $\nabla V=\nabla {{V}_{1}}+\nabla {{V}_{2}}$, Therefore
\begin{equation}
    \begin{aligned}
&V(t+1) - V(t) \\
&\le -\left[ ({s}_{1} - \beta {s}_{2})\left\Vert{q}\right\Vert^2 - \left( \frac{{s}_{2}}{\beta} + {s}_{3} \right)\left\Vert{e}\right\Vert^2 \right] - {{\text{W}}_{2}}{{\delta }^{T}}\delta\\
& - (-{{\rho }_{i}}){{\mu }_{i}}(t) + \vartheta \left( \frac{\partial({{s}_{1}} - {{s}_{2}}\beta)}{({{s}_{2}} + {{s}_{3}}\beta){{\beta }^{-1}}}\left\Vert{{q}_{l}}\right\Vert^2 - \left\Vert{{e}_{l}}\right\Vert^2 \right) \\
& \le \left( {{s}_{2}}{{\beta }^{-1}} + {{s}_{3}} - \vartheta \right)\left\Vert{e}\right\Vert^2 
+ (-{{\rho }_{i}}){{\mu }_{i}}(k) - {{\text{W}}_{2}}{{\delta }^{T}}\delta -\\
& \left( \frac{(+\partial({{s}_{2}}{{\beta }^{-1}} + {{s}_{3}}) - \partial({{s}_{2}}{{\beta }^{-1}} + {{s}_{3}})) + ({{s}_{2}}{{\beta }^{-1}} + {{s}_{3}})}{\partial} - \vartheta \right) \\
&\times \frac{\partial({{s}_{1}} - \beta {{s}_{2}})}{({{s}_{2}}{{\beta }^{-1}} + {{s}_{3}})}\left\Vert{q}\right\Vert^2 \le \left( {{s}_{2}}{{\beta }^{-1}} + {{s}_{3}} - \vartheta \right)\left\Vert{e}\right\Vert^2\\
& - \left( {{s}_{2}}{{\beta }^{-1}} + {{s}_{3}} - \vartheta \right)\frac{\partial({{s}_{1}} - \beta {{s}_{2}})}{({{s}_{2}}{{\beta }^{-1}} + {{s}_{3}})}\left\Vert{q}\right\Vert^2\\
&+ (-{{\rho }_{i}}){{\mu }_{i}}(k) - {{\text{W}}_{2}}{{\delta }^{T}}\delta \le -\left[ \left( 1 - \partial \right)\left( {{s}_{1}} - \beta {{s}_{2}} \right) \right]\left\Vert{q}\right\Vert^2 \\
&- {{\text{W}}_{2}}{{\delta }^{T}}\delta + \left( -{{\rho }_{i}} \right){{\mu }_{i}}(k) + \left( {{s}_{2}}{{\beta }^{-1}} + {{s}_{3}} - \vartheta \right)\frac{{{\mu }_{i}}}{{{\theta }_{i}}}
 \label{eq_68}
    \end{aligned}
\end{equation}

Similar to the previous section, by converting equation (\ref{eq_33}) to (\ref{eq_42}), one gets

\[
\begin{aligned}
&\quad V\left( t+1 \right)-V\left( t \right)
\le  \notag \\
\end{aligned}
\]
\[
\begin{aligned}
&\quad-[\left( 1-\partial  \right)~({{s}_{1}}-\beta {{s}_{2}})
\times {{\left( {{\lambda }_{N}}^{-1}-\sqrt{\frac{\partial \left( {{s}_{1}}-{{s}_{2}}\beta  \right)}{({{s}_{2}}+{{s}_{3}}\beta ){{\beta }^{-1}}}} \right)}^{-2}}   \notag 
\end{aligned}
\]
\begin{equation}
 +{{\text{W}}_{2}}{\left\Vert{\delta}\right\Vert^2}
-\left( +{{\rho }_{i}}-\frac{{{s}_{2}}{{\beta }^{-1}}+{{s}_{3}}-\vartheta }{{{\theta }_{i}}} \right){{\mu }_{i}} 
\label{eq_70}
\end{equation}

When the platoon system is free of attack, that is,
$~t\in \left[ {{T}_{k}},{{T}_{k+1}} \right]~~,~{{T}_{k}}\notin \Xi \left( {{t}_{0}}~,~t \right)$ one has

\[
\begin{aligned}
&\quad V\left( t+1 \right)-V\left( t \right) \le \notag
\end{aligned}
\]
\[
\begin{aligned}
 -[\left( 1-\partial  \right)~({{s}_{1}}-\beta {{s}_{2}})  {{\left( {{\lambda }_{N}}^{-1}-\sqrt{\frac{\partial \left( {{s}_{1}}-{{s}_{2}}\beta  \right)}{({{s}_{2}}+{{s}_{3}}\beta ){{\beta }^{-1}}}} \right)}^{-2}} +{{\text{W}}_{2}}]~\left\Vert{\delta}\right\Vert^2\notag \\
\end{aligned}
\]
\[
\begin{aligned}
&\quad  -\left( +{{\rho }_{i}}-\frac{{{s}_{2}}{{\beta }^{-1}}+{{s}_{3}}-\vartheta }{{{\theta }_{i}}} \right){{\mu }_{i}}~ \notag 
\end{aligned}
\]

\begin{equation}
\le -{{\alpha }_{1}}{{V}_{1}}-{{\alpha }_{2}}{{V}_{2}}\le -\tilde{\Gamma}V  
\label{eq_71}
\end{equation}

Therefore
\begin{equation}
    \begin{aligned}
V\left( t \right)<{{e}^{-~\tilde{\Gamma}\left( t-{{T}_{n}} \right)}}V\left( {{T}_{n}} \right),t\in \left[ {{T}_{k}},{{T}_{k+1}} \right],{{T}_{k}}\notin \Xi \left( {{t}_{0}}~,~t \right)
 \label{eq_72}
    \end{aligned}
\end{equation}
Due to $\tilde{\Gamma}$ being a positive number. Thus, it follows that $V\left( t \right)\overset{~}{\mathop{\to }}\,0$ as t$\overset{~}{\mathop{\to }}\,\infty $. This implies that platoon system consensus can be asymptotically achieved.
When attack is activated, $t\in \left[ {{T}_{k}},{{T}_{k+1}} \right]~~,~{{T}_{k}}\in \Xi \left( {{t}_{0}}~,~t \right)$, it is obtained that

\[
\begin{aligned}
&\quad V\left( t+1 \right)-V\left( t \right) \le \notag 
\end{aligned}
\]
\[
\begin{aligned}
&\quad -[\left( 1-\partial  \right)~({{s}_{1}}-\beta {{\tilde{s}}_{2}}) {{\left( {{\lambda }_{N}}^{-1}-\sqrt{\frac{\partial \left( {{s}_{1}}-{{{\tilde{s}}}_{2}}\beta  \right)}{({{{\tilde{s}}}_{2}}+{{{\tilde{s}}}_{3}}\beta ){{\beta }^{-1}}}} \right)}^{-2}} \notag 
\end{aligned}
\]
\[
\begin{aligned}
&\quad +{{\tilde{W}}_{2}}]~\left\Vert{\delta}\right\Vert^2 -\left( {{\rho }_{i}}-\frac{{{{\tilde{s}}}_{2}}{{\beta }^{-1}}+{{{\tilde{s}}}_{3}}-\vartheta }{{{\theta }_{i}}} \right){{\mu }_{i}} \notag \\
\end{aligned}
\]

\begin{equation}
\le \tilde{\gamma}{{V}_{1}}-{{\alpha }_{2}}{{V}_{2}}\le \tilde{\gamma}V 
\label{eq_73}
\end{equation}

Then, one obtains
\begin{equation}
    \begin{aligned}
V~\left( t \right)<{{e}^{-~\tilde{\gamma}\left( t-{{T}_{{{k}'}}} \right)}}V\left( {{T}_{n}} \right),~t\in \left[ {{T}_{{{k}'}}},{{T}_{{k}'+1}} \right],{{T}_{{{k}'}}}\in \Xi \left( {{t}_{0}}~,~t \right)
 \label{eq_74}
    \end{aligned}
\end{equation}
Similar to Theorem 1, it can be derived that
\begin{equation}
    \begin{aligned}
V\left( t \right)\le V\left( {{t}_{0}} \right)~{{e}^{\left( \tilde{\Gamma}+\tilde{\gamma} \right)\left( {{\zeta }_{0}}+T \right)}}~{{e}^{\left[ -\tilde{\Gamma}+~\left( \tilde{\Gamma}+\tilde{\gamma} \right){{\tau }_{0}} \right]\left(t-{{t}_{0}} \right)}}
 \label{eq_75}
    \end{aligned}
\end{equation}

According to (\ref{eq_64}), we can conclude that the asymptotic consensus of the platoon (\ref{eq_1}) is achieved. The proof is finished.

\emph{Remark 4}: If triggering parameters ${{\rho }_{i}}$ $,\frac{\vartheta }{{{\theta }_{i}}}~$in dynamic auxiliary variable$~{{\mu }_{i}}$ are chosen to be relatively small, the left term in (\ref{eq_62}) is not larger than $\tilde{\alpha }$. As a result, the convergence rate of the system becomes slower compared to a static ETC. This observation suggests that although the dynamic protocol can reduce the frequency of triggering instants compared to the static protocol, it comes at the cost of degraded system performance. The tradeoff between system performance and triggering frequency is verified in Theorems 1 and 2.

\section{Attack mitigation}
In scenarios where there are no constraints on attack frequency and duration, the guarantees for secure consensus, as mentioned in the previous section, cannot be applied. In such cases, alternative approaches for attack mitigation are necessary. Soodeh Dadras et al. proposed a control algorithm based on fractional order principles to prevent collisions in an adversarial platooning environment \cite{dadras2019resilient}. Another approach demonstrated in \cite{khanapuri2021learning}, involves adjusting the $k_v$ values of non-attacker vehicles to mitigate attacks. However, it's crucial to consider the worst-case scenario where attackers can affect the $k_v$ values in all vehicles. To prevent attackers from inducing collisions in a vehicular platoon, the proposed approach involves extending the allowable bounds of control gains. By deriving sufficient conditions, it becomes possible to increase the upper bound of $k_v$ to mitigate attacks and stabilize the platoon. Equation (\ref{eq_17}) illustrates that raising the upper bound of $k_v$ can effectively mitigate attacks. To achieve this, either ${{\lambda }{\text{max}}}$ needs to decrease or $k_p$ should increase while satisfying condition (\ref{eq_17}) under attack conditions. Due to constraints on accessing the controller input, attacks can be mitigated by reducing ${{\lambda }{\text{max}}}$. This entails selecting a new network topology for the platoon system under attack, where the new topology has a lower ${{\lambda }_{\text{max}}}$ that satisfies stability conditions. Choosing a new network topology that meets these conditions effectively prevents attackers from inducing collisions.

\emph{Remark 5}: In selecting a new topology based on the lowest $~{{\lambda }_{\text{max}}}$ value, it may not be possible to find a single topology that consistently has the lowest $~{{\lambda }_{\text{max}}}$ value across all communication ranges. It is important to consider performance criteria to limit the candidate topologies, for example, the limitation of communication bandwidth in a hierarchical platoon network.

\begin{figure}[ht!] 
\centering
\includegraphics[width=2.8in]{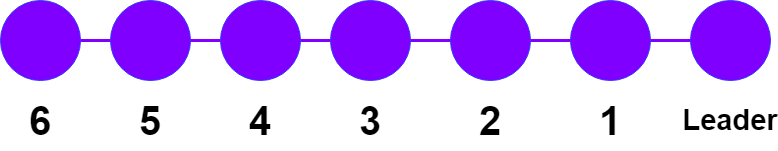}
\caption{  Communication topology of the platoon. }
\label{fig_1}
\end{figure}

\section{Simulation}
This section presents numerical simulations to validate the derived results. Consider a platoon with a leader (\ref{eq_2}) and six following vehicles (\ref{eq_1}). The sampling period is set to $\hat{T}$ = 0.2 seconds. The platoon adopts a bidirectional (BD) topology, as shown in Fig. \ref{fig_1}. The adjacency matrix $H$ and $\bar{\L}$ can be found in Appendix B. The initial states of vehicles are as follows: ${{x}_{0}}={{\left[ 100,12 \right]}^{T}}$, ${{x}_{1}}={{\left[ 65,10 \right]}^{T}}$, ${{x}_{2}}={{\left[ 40,8 \right]}^{T}}$, ${{x}_{3}}={{\left[11,6 \right]}^{T}}$, ${{x}_{4}}={{\left[ 0,4 \right]}^{T}}$, ${{x}_{5}}={{\left[-20,2 \right]}^{T}}$, ${{x}_{6}}={{\left[ -25,0 \right]}^{T}}$.
 From the communication topology, we can  calculate ${{\lambda }_{1}}=0.058$ and$~{{\lambda }_{N}}=3.77$. As a result, conditions (\ref{eq_13}) and (\ref{eq_15}) hold, respectively. Additionally, we have $\underset{j}{\mathop \prod }\,\left| \lambda _{j}^{u}\left( A \right) \right|=1<1.0313~~,$  and $~~0.04<{{\lambda }_{i}}<3.94$, for $i=1,2,\ldots ,N$. We can calculate the controller gain from (\ref{eq_16}) as $k=\left[ 0.1259~2.5252 \right]$ then according to \cite{yan2020analysis}, the platoon achieves consensus. The desired spacing between two consecutive vehicles is set as $\nabla =20m$. As it turns out, The control gain falls within the stability range of equation (\ref{eq_17}), i.e., $0~<\hat{T}{{k}_{p}}=0.025<{{k}_{v}}=2.52<\frac{2}{\hat{T}{{\lambda}_{\text{max}}}}+\frac{\hat{T}{{k}_{p}}}{2}=2.67$. Thus, a successful attack would require increasing ${{k}_{v}}~$ to more than 2.68 in order to destabilize the platoon.
 In simulation example 1, we discuss secure event-triggered control strategies under the sequential gain modification attack in two cases. The assumption is made that assumptions 2 and 3 are satisfied. In example 2, a different approach is introduced to resist the attacker when no constraint is imposed on the duration and frequency of the attack. The purpose of both examples is to explore different strategies to secure event-triggered control in the presence of attacks and to evaluate their effectiveness under different assumptions and attack scenarios.

\begin{figure}[ht!] 
\centering
\includegraphics[ width=3.45in]{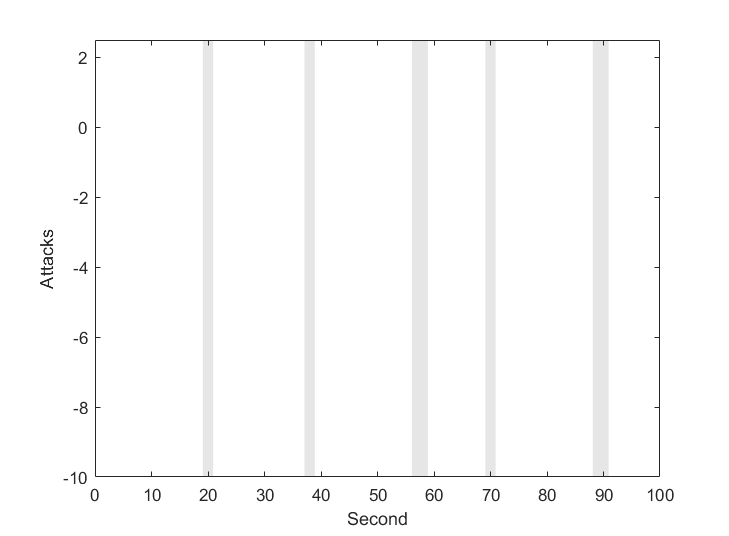}
\caption{Attack sequence of Example 1.}
\label{fig_5}
\end{figure}

\emph{Example 1}:
 We consider a randomly launched gain modification attack sequence. In two cases, we investigate a secure event-triggered consensus control system where a malicious adversary attacks the controller gain of some or all vehicles. The attack instants are shown in Fig. \ref{fig_5}. The attack sequence satisfies $~{{\zeta }_{0}}=0.12~,~{{\tau }_{0}}=3~~,~{{\mathcal{F}}_{0}}=4~$and$~{{f}_{0}}=0.05$.

\emph{Case 1 (Static Event-Triggered Control)}: 
 For the static event-triggered control protocol, we set the variable$~\partial \left( t \right)=0.01$. Using equation (\ref{eq_16}) we calculate the controller gain to be $k=\left[ 0.1259~2.5252 \right]$ and the parameter $~~\tilde{\alpha }=0.0782$. 
 
When the platoon system is under attack and the controller gain increased by ${{\tilde{g}}_{v}}=0.58$, we can calculate ${k_{g}}=\left[ 0.1259\text{ }\!\!~\!\!\text{  }\!\!~\!\!\text{ },3.25 \right]$ which leads to system instability. Additionally, we obtain $~\tilde{\gamma}=0.3414$. Therefore, $\frac{{\tilde{\alpha}}}{\tilde{\alpha}+\tilde{\gamma}}=0,1864$. Using simulation, we determine that ${{\Delta}^{*}}=0.4$. Based on Theorem 1, ${{\tau }_{0}}+{{\Delta}^{*}}{{f}_{0}}=0.14<0.1864~$ which satisfies the condition (\ref{eq_25}). Time evolutions of consensus errors are depicted in Fig. \ref{fig_6}, implying that secure consensus is achieved asymptotically. In this static scheme, the average number of triggered events per vehicle is 278.8 times in 500 steps.



\begin{figure*}[!t]
\centering
\subfloat[\centering ]{\includegraphics[width=7.5cm]{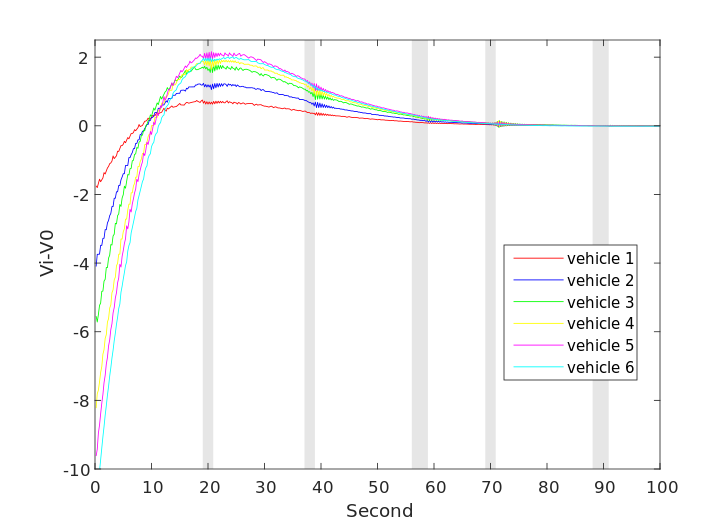}%
\label{fig_6}}
\hfil
\subfloat[\centering ]{\includegraphics[width=7.5cm]{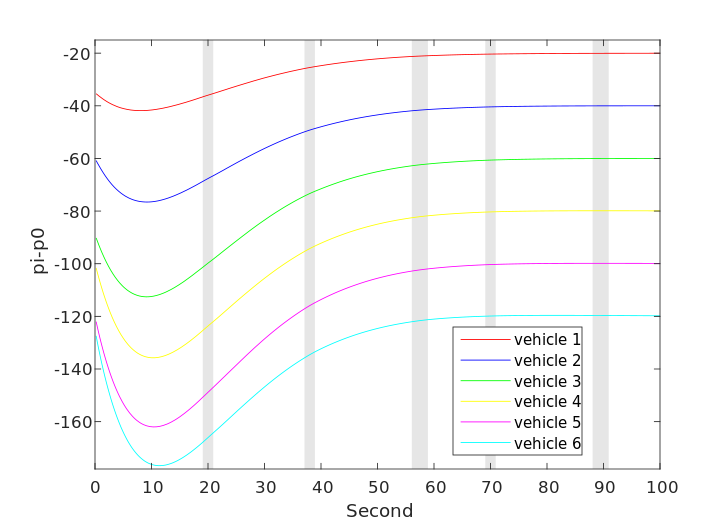}%
\label{fig_6}}
\caption{Time evolutions of platoon consensus under the static event-triggered scheme. (a) Velocity errors. (b) Spacing errors.}
\label{fig_6}
\end{figure*}

\begin{figure}%
    \centering
    \subfloat[\centering ]{{\includegraphics[width=7.5cm]{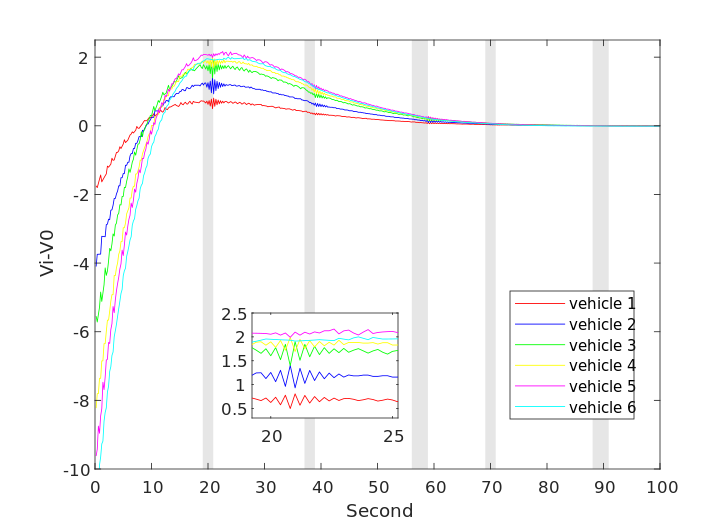} }}%
    \qquad
    \subfloat[\centering ]{{\includegraphics[width=7.5cm]{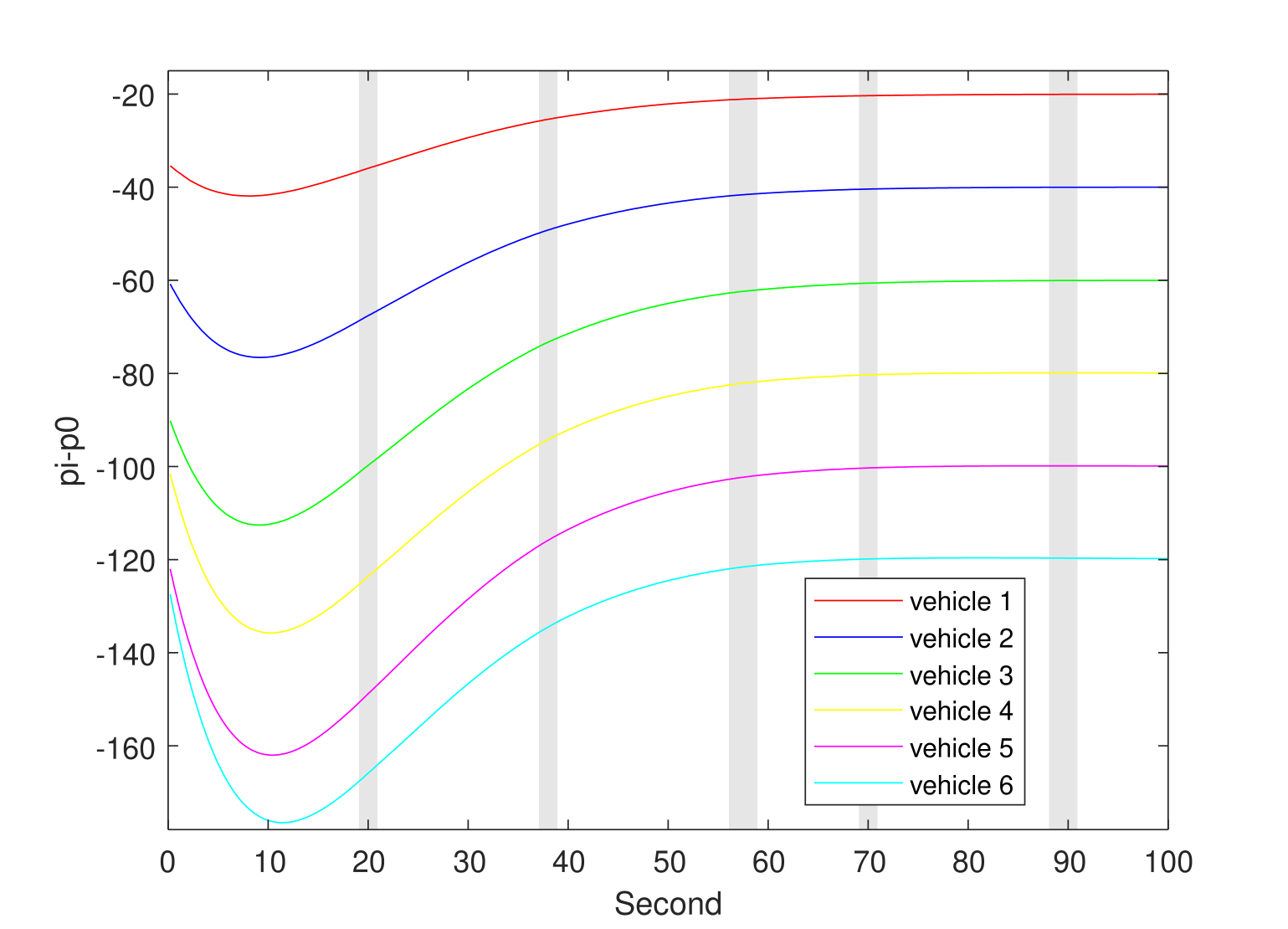} }}%
    \caption{Time evolutions of platoon consensus under the dynamic event-triggered scheme. (a) Velocity errors (b) Spacing errors.}%
    \label{fig_7}%
\end{figure}

\begin{figure}[ht!] 
\centering
\includegraphics[width=3.45in]{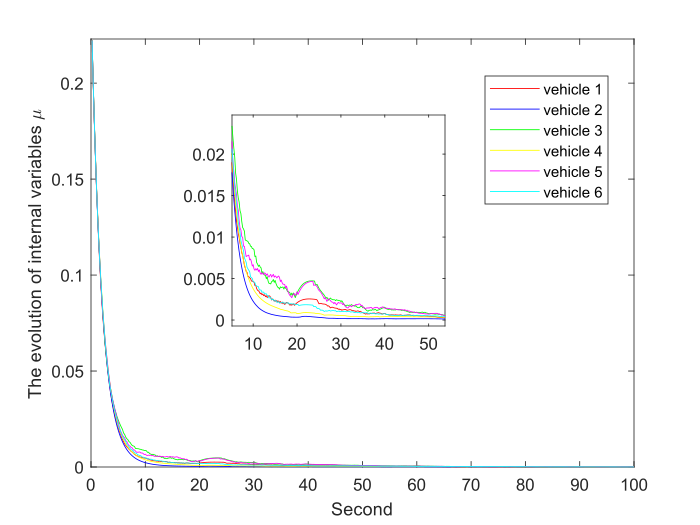}
\caption{Thresholds for vehicles under the dynamic event-triggering law. }
\label{fig_8}
\end{figure}

\emph{Case 2 (Dynamic Event-Triggered Control)}: 
Under similar conditions as in Case 1, we have $\tilde{\alpha }=0.0782~$ and $\tilde{\gamma}=0.3414~$. As mentioned in Remark 4, ${{\alpha }_{1}}<\tilde{\alpha}$ provides evidence of the superiority of the dynamic event-triggered scheme with a lower triggering frequency.
By choosing the design parameters $~{{\rho }_{i}}=0.1,~~{{\vartheta }_{i}}=~0.6~,{{\theta }_{i}}=90~~$ and initialize ${{\mu }_{i}}\left( 0 \right)=20$. According to (\ref{eq_61}) and (\ref{eq_62}) in Theorem 2, one derive ${{\alpha }_{1}}=\tilde{\Gamma}=0.0750<\tilde{\alpha }=0.0782$ and$~\frac{\tilde{\Gamma}}{\tilde{\Gamma}+\tilde{\gamma}}=0.1801$. Subsequent simulation reveals $~{{\Delta}^{*}}=0.8$ thereby leading to $~{{\tau }_{0}}+{{\Delta}^{*}}{{f}_{0}}=0.16<\frac{\tilde{\Gamma}}{\tilde{\Gamma}+\tilde{\gamma}}=0.1801$, satisfying (\ref{eq_64}). Consequently, in accordance with Theorem 2, we ascertain the feasibility of achieving secure consensus. Fig. \ref{fig_7} illustrates the consensus errors of the vehicles in the dynamic case and is accompanied by the corresponding triggering instants of each vehicle showing that the average number of triggered events per vehicle for the dynamic scheme is 221.1 times in 500 simulation steps. Figure \ref{fig_8} illustrates the behavior of the threshold variable ${{\mu }_{i}}\left( t \right)$ for each vehicle in the dynamic triggering law. These thresholds converge to zero over time, indicating that the triggering instants decrease as the system approaches consensus, but
sometimes it may be an increase.


By setting the error threshold  $E\left( t \right)$ = max $\underset{t\overset{~}{\mathop{\to }}\,\infty }{\mathop{\lim }}\,\left| {{v}_{i}}\left( t \right)-{{v}_{0}}\left( t \right) \right|$ to $10^{-2}$ as an index for achieving consensus in the vehicle platoon system. In the static scheme (case 1), consensus is reached at $t = 86.2s$, while in the dynamic scheme (case 2), consensus is reached at $t = 87.2s$. These consensus times satisfy the actual requirements of the platoon system based on the consensus index. The index $J = ([\underset{i\in {{N}_{i}}}{\mathop \sum } Q_{i}]/Ntc)$ is defined as the average triggering rate, where the number of triggering times $Q_{i}$ is triggered to achieve consensus for vehicle $i$. $N$ is the total number of vehicles, and $t_{c}$ represents the achieved consensus time. In the static scheme, the value of $J_{static}$ is 3.23, whereas $J_{dynamic}$ is 2.53 for the dynamic scheme. These average triggering rates provide a measure of the frequency at which the vehicles in the platoon need to communicate and update their states to reach a consensus. Table I provides the comparison of triggering instants between static and dynamic schemes when consensus is achieved. Evidently, the dynamic event-triggered controller decreases the number of triggering instants compared to the static scheme. However, this reduction in triggering instants comes at the expense of consensus performance, indicating that the dynamic event-triggered control scheme may achieve consensus at a slower rate or with less accuracy than the static scheme.

\begin{table}[h!]
\caption{Comparisons of triggering times by different
Event-triggered mechanisms under attack. }
\centering
\begin{tabular}{ |p{1cm}|p{0.5cm}|p{0.5cm}|p{0.5cm}|p{0.5cm}|p{0.5cm}|p{0.5cm}|p{0.7cm}|  }
\hline
\multicolumn{8}{|c|}{Triggering numbers for Vehicles} \\
\hline
Scheme & 1 &2&3&4&5&6&Total \\
\hline
Static & 198 &266&194&282&227&248&1415 \\
\hline
Dynamic & 163 &242&146&244&172&206&1173 \\
\hline
\end{tabular}
\label{table:tabale_name}
\end{table}

\begin{figure}[ht!] 
\centering
\includegraphics[width=2.8in]{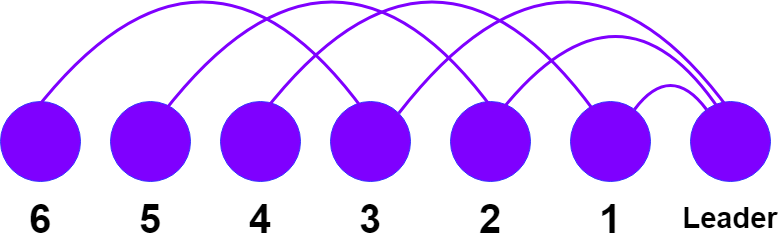}
\caption{Information flow topology of all vehicles.}
\label{fig_9}
\end{figure}

\begin{figure}%
    \centering
    \subfloat[\centering ]{{\includegraphics[width=7.5cm]{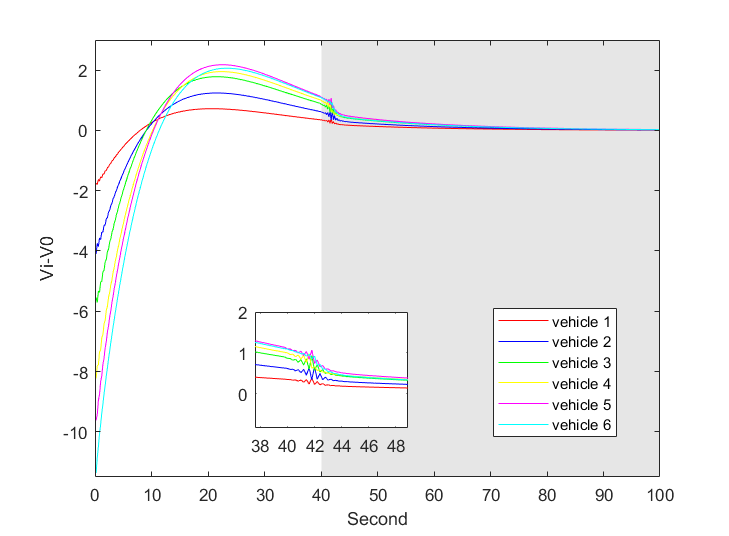} }}%
    \qquad
    \subfloat[\centering ]{{\includegraphics[width=7.5cm]{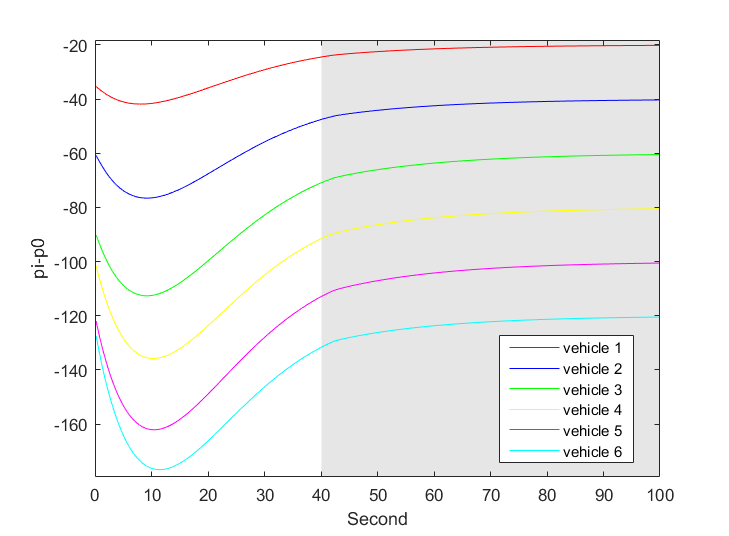} }}%
    \caption{States evolution for secure platoon consensus in the
presence of modification attacks. (a) Velocity errors (b) Spacing errors.}%
    \label{fig_10}%
\end{figure}

\emph{Example 2}: In the given scenario, it is mentioned that the adversaries have no constraint on the duration and frequency of the attack. According to equation (\ref{eq_17}), reducing the maximum eigenvalue (${{\lambda }_{\text{max}}}$) will increase the upper bounds of ${{k}_{v}}$. It is also mentioned that under $k=\left[ 0.1259\text{ }\!\!~\!\!\text{  }\!\!~\!\!\text{ }3.25 \right]$, the platoon system becomes unstable because when $~{{k}_{v}}>2.68~$ the platoon is unstable. In Fig. \ref{fig_9}, a topology with fewer $~{{\lambda }_{\text{max}}}$ than the BD topology is chosen to resist this attack. The impact of this topology on attack mitigation is shown in Fig. \ref{fig_10}. 

Fig. \ref{fig_10} illustrates the simulation result for the platoon system, the same as example 1, which remains stable until $t=40s$. However, a gain modification attack occurred in the control gain after this time, leading to a violation of assumptions 2 and 3. To counter this attack, the system switched to a different topology at the 43rd second, as shown in Fig. \ref{fig_9}. By employing this new topology at ${t}_{s}=43$ under $k=\left[ 0.1259\text{ }\!\!~\!\!\text{  }\!\!~\!\!\text{ }, 3.25 \right]$ , the platoon system remains stable under attack. This is confirmed, which yields $0~<{{g}_{v}}+{{\tilde{g}}_{v}}=3.25<\frac{2}{\hat{T}{{\lambda }_{\text{max}}}}+\frac{\hat{T}{{g}_{s}}}{2}=3.85$.  Consequently, as the value of ${{\tilde{g}}_{v}}$ increases, a smaller value of $~{{\lambda }_{\text{max}}}$ is required to mitigate the attack. Hence, it is crucial to adapt the platoon system's topology and parameters according to the severity of the attack in order to maintain stability.

\section{Conclusion}
This article has addressed the event-triggered secure consensus problem for second-order vehicle platooning in the discrete-time domain, considering a gain modification attack. Static and dynamic event-triggering strategies have been proposed to achieve secure consensus. we derived sufficient conditions that incorporate constraints on the attack duration and frequency. These conditions are related to the triggering parameters, maximum, and minimum eigenvalues of the network topology and platoon system matrices. Also, conditions based on the Schur stability criterion have been applied to defend against gain modification attacks without constraints on duration and frequency. Simulations have been conducted to validate the effectiveness of the developed methods.

\section{Appendix}
\subsection{Proof of Lemma 1}
\emph{Proof}: The closed-loop dynamics of the vehicle platooning system can be rewritten as

\[
\begin{aligned}
&\quad   x\left( t+1 \right)=P\text{ }\!\!~\!\!\text{ }x\left( t \right)  \notag \\
\end{aligned}
\]

\begin{equation}
 P={{I}_{N-1}}\otimes \text{A}-{{L}_{g}}\otimes BK,\text{ }\!\!~\!\!\text{  }\!\!~\!\!\text{  }\!\!~\!\!\text{  }\!\!~\!\!\text{ }BK=\left[ \begin{matrix}
   0 & 0  \\
   \hat{T}{{k}_{p}} & \hat{T}{{k}_{v}}  \\
\end{matrix} \right] 
\end{equation}
The stability of x(t) in (\ref{eq_15}) depends on ${\bold{P}}$ and the spectrum of P is:
\begin{equation}
    \begin{aligned}
 \sigma \left( P \right)= \underset{{{\lambda }_{l}}\in \sigma \left( {H} \right)}{\medcup} \left\{ \sigma \left( A-{{\lambda }_{l}}BK \right) \right\}\\
	=\underset{{{\lambda }_{l}}\in \sigma \left( {H} \right)}{\medcup}\,\left\{ \sigma \left( \begin{matrix}
   1 & \hat{T}  \\
   -{{\lambda }_{l}}\hat{T}{{k}_{p}} & 1-\hat{T}{{k}_{v}}  \\
\end{matrix}  \right) \right\}=Q
   \end{aligned}
\end{equation}
Where$~\sigma \left( . \right)$ is the set of distinct eigenvalues. The eigenvalues of Q, denoted by Z, are the roots of the following equation
\begin{equation}
    \begin{aligned}
\omega \left( Z \right)={{Z}^{2}}+\left( {{\lambda }_{l}}\hat{T}{{k}_{v}}-2 \right)Z+{{\lambda }_{l}}{{\hat{T}}^{2}}{{k}_{p}}-{{\lambda }_{l}}\hat{T}{{k}_{v}}+1=0
    \end{aligned}
\end{equation}
From (\ref{eq_22}), we know that all eigenvalues of ${{L}_{g}}$ are positive. As  ${{\lambda }_{l}}>0$~,~l=1,2,\ldots, N, Z falls into the unit disk, i.e., $\left| Z \right|<1$. Using bilinear transformation, we can determine the Schur stability of $\omega \left( Z \right)$ by determining the Hurwitz stability of the corresponding continuous-time system. So $P$ is Schur stable.
\subsection{The topologies matrix of the platoon} 
 \[
\bar{L}_{\text{BD}} = \begin{bmatrix}
     1 & -1 & 0 & 0 & 0 & 0 & 0 \\
     -1 & 2 & -1 & 0 & 0 & 0 & 0\\
     0 & -1 & 2 & -1 & 0 & 0 & 0\\
     0 & 0 & -1 & 2 & -1 & 0 & 0\\
     0 & 0 & 0 & -1 & 2 & -1 & 0\\
     0 & 0 & 0 & 0 & -1 & 2 & -1\\  
     0 & 0 & 0 & 0 & 0 & -1 & 1\\  
 \end{bmatrix}
\bar{L}_{\text{Switched}} = \begin{bmatrix}
 3 & -1 & -1 & -1 & 0 & 0 & 0 \\
 -1 & 2 & 0 & 0 & -1 & 0 & 0 \\
 -1 & 0 & 2 & 0 & 0 & -1 & 0 \\
 -1 & 0 & 0 & 2 & 0 & 0 & -1 \\
 0 & -1 & 0 & 0 & 1 & 0 & 0 \\
 0 & 0 & -1 & 0 & 0 & 1 & 0 \\
 0 & 0 & 0 & -1 & 0 & 0 & 1 \\
\end{bmatrix}
\]
  
$H$ is obtained for each matrix by deleting the first row and column of $\bar{L}$.

\bibliographystyle{IEEEtran}  
\bibliography{references.bib}

\begin{thebibliography}{10}
\providecommand{\url}[1]{#1}
\csname url@samestyle\endcsname
\providecommand{\newblock}{\relax}
\providecommand{\bibinfo}[2]{#2}
\providecommand{\BIBentrySTDinterwordspacing}{\spaceskip=0pt\relax}
\providecommand{\BIBentryALTinterwordstretchfactor}{4}
\providecommand{\BIBentryALTinterwordspacing}{\spaceskip=\fontdimen2\font plus
\BIBentryALTinterwordstretchfactor\fontdimen3\font minus \fontdimen4\font\relax}
\providecommand{\BIBforeignlanguage}[2]{{%
\expandafter\ifx\csname l@#1\endcsname\relax
\typeout{** WARNING: IEEEtran.bst: No hyphenation pattern has been}%
\typeout{** loaded for the language `#1'. Using the pattern for}%
\typeout{** the default language instead.}%
\else
\language=\csname l@#1\endcsname
\fi
#2}}
\providecommand{\BIBdecl}{\relax}
\BIBdecl

\bibitem{chen2010review}
B.~Chen and H.~H. Cheng, ``A review of the applications of agent technology in traffic and transportation systems,'' \emph{IEEE Transactions on intelligent transportation systems}, vol.~11, no.~2, pp. 485--497, 2010.

\bibitem{caveney2010cooperative}
D.~Caveney, ``Cooperative vehicular safety applications,'' \emph{IEEE Control Systems Magazine}, vol.~30, no.~4, pp. 38--53, 2010.

\bibitem{zheng2015stability}
Y.~Zheng, S.~E. Li, J.~Wang, D.~Cao, and K.~Li, ``Stability and scalability of homogeneous vehicular platoon: Study on the influence of information flow topologies,'' \emph{IEEE Transactions on intelligent transportation systems}, vol.~17, no.~1, pp. 14--26, 2015.

\bibitem{you2011network}
K.~You and L.~Xie, ``Network topology and communication data rate for consensusability of discrete-time multi-agent systems,'' \emph{IEEE Transactions on Automatic Control}, vol.~56, no.~10, pp. 2262--2275, 2011.

\bibitem{yan2020analysis}
Y.~Yan, S.~Stuedli, M.~M. Seron, and R.~H. Middleton, ``Analysis of attack via grounding and countermeasures in discrete-time consensus networks,'' \emph{arXiv preprint arXiv:2002.11938}, 2020.

\bibitem{wang2020adaptive}
J.~Wang, F.~Ma, Y.~Yang, J.~Nie, B.~Aksun-Guvenc, and L.~Guvenc, ``Adaptive event-triggered platoon control under unreliable communication links,'' \emph{IEEE Transactions on Intelligent Transportation Systems}, vol.~23, no.~3, pp. 1924--1935, 2020.

\bibitem{zhang2020distributeds}
H.~Zhang, J.~Liu, Z.~Wang, H.~Yan, and C.~Zhang, ``Distributed adaptive event-triggered control and stability analysis for vehicular platoon,'' \emph{IEEE Transactions on Intelligent Transportation Systems}, vol.~22, no.~3, pp. 1627--1638, 2020.

\bibitem{nowzari2019event}
C.~Nowzari, E.~Garcia, and J.~Cort{\'e}s, ``Event-triggered communication and control of networked systems for multi-agent consensus,'' \emph{Automatica}, vol. 105, pp. 1--27, 2019.

\bibitem{Tabuada2007real-time}
P.~Tabuada, ``Event-triggered real-time scheduling of stabilizing control tasks,'' \emph{IEEE Transactions on Automatic control}, vol.~52, no.~9, pp. 1680--1685, 2007.

\bibitem{ding2017overview}
L.~Ding, Q.-L. Han, X.~Ge, and X.-M. Zhang, ``An overview of recent advances in event-triggered consensus of multiagent systems,'' \emph{IEEE transactions on cybernetics}, vol.~48, no.~4, pp. 1110--1123, 2017.

\bibitem{liu2020event}
J.~Liu, T.~Yin, D.~Yue, H.~R. Karimi, and J.~Cao, ``Event-based secure leader-following consensus control for multiagent systems with multiple cyber attacks,'' \emph{IEEE Transactions on Cybernetics}, vol.~51, no.~1, pp. 162--173, 2020.

\bibitem{dimarogonas2009event}
D.~V. Dimarogonas and K.~H. Johansson, ``Event-triggered control for multi-agent systems,'' in \emph{Proceedings of the 48h IEEE Conference on Decision and Control (CDC) held jointly with 2009 28th Chinese Control Conference}.\hskip 1em plus 0.5em minus 0.4em\relax IEEE, 2009, pp. 7131--7136.

\bibitem{wu2017event}
Z.-G. Wu, Y.~Xu, R.~Lu, Y.~Wu, and T.~Huang, ``Event-triggered control for consensus of multiagent systems with fixed/switching topologies,'' \emph{IEEE Transactions on Systems, Man, and Cybernetics: Systems}, vol.~48, no.~10, pp. 1736--1746, 2017.

\bibitem{xu2015event}
W.~Xu, D.~W. Ho, L.~Li, and J.~Cao, ``Event-triggered schemes on leader-following consensus of general linear multiagent systems under different topologies,'' \emph{IEEE transactions on cybernetics}, vol.~47, no.~1, pp. 212--223, 2015.

\bibitem{liu2019cooperative}
Z.~Liu, Z.~Li, G.~Guo, and H.~Cheng, ``Cooperative platoon control of heterogeneous vehicles under a novel event-triggered communication strategy,'' \emph{IEEE Access}, vol.~7, pp. 41\,172--41\,182, 2019.

\bibitem{linsenmayer2015event}
S.~Linsenmayer and D.~V. Dimarogonas, ``Event-triggered control for vehicle platooning,'' in \emph{2015 American control conference (ACC)}.\hskip 1em plus 0.5em minus 0.4em\relax IEEE, 2015, pp. 3101--3106.

\bibitem{li2019string}
Z.~Li, B.~Hu, M.~Li, and G.~Luo, ``String stability analysis for vehicle platooning under unreliable communication links with event-triggered strategy,'' \emph{IEEE Transactions on Vehicular Technology}, vol.~68, no.~3, pp. 2152--2164, 2019.

\bibitem{wen2018event}
S.~Wen, G.~Guo, B.~Chen, and X.~Gao, ``Event-triggered cooperative control of vehicle platoons in vehicular ad hoc networks,'' \emph{Information Sciences}, vol. 459, pp. 341--353, 2018.

\bibitem{ge2021dynamic}
X.~Ge, S.~Xiao, Q.-L. Han, X.-M. Zhang, and D.~Ding, ``Dynamic event-triggered scheduling and platooning control co-design for automated vehicles over vehicular ad-hoc networks,'' \emph{IEEE/CAA Journal of Automatica Sinica}, vol.~9, no.~1, pp. 31--46, 2021.

\bibitem{ge2020dynamic}
X.~Ge, Q.-L. Han, L.~Ding, Y.-L. Wang, and X.-M. Zhang, ``Dynamic event-triggered distributed coordination control and its applications: A survey of trends and techniques,'' \emph{IEEE Transactions on Systems, Man, and Cybernetics: Systems}, vol.~50, no.~9, pp. 3112--3125, 2020.

\bibitem{he2019adaptive}
W.~He, B.~Xu, Q.-L. Han, and F.~Qian, ``Adaptive consensus control of linear multiagent systems with dynamic event-triggered strategies,'' \emph{IEEE Transactions on Cybernetics}, vol.~50, no.~7, pp. 2996--3008, 2019.

\bibitem{xiao2021dynamic}
S.~Xiao, X.~Ge, Q.-L. Han, and Y.~Zhang, ``Dynamic event-triggered platooning control of automated vehicles under random communication topologies and various spacing policies,'' \emph{IEEE Transactions on Cybernetics}, vol.~52, no.~11, pp. 11\,477--11\,490, 2021.

\bibitem{karaki2021distributed}
B.~J. Karaki and M.~S. Mahmoud, ``Distributed event-triggered consensus protocols for discrete-time multiagent systems,'' \emph{IMA Journal of Mathematical Control and Information}, vol.~38, no.~4, pp. 1046--1071, 2021.

\bibitem{mishra2021dynamic}
R.~K. Mishra and H.~Ishii, ``Dynamic event-triggered consensus control of discrete-time linear multi-agent systems,'' \emph{IFAC-PapersOnLine}, vol.~54, no.~17, pp. 123--128, 2021.

\bibitem{mishra2021average}
------, ``Average consensus in discrete-time multi-agent systems with distributed event-triggered control,'' in \emph{2021 European Control Conference (ECC)}.\hskip 1em plus 0.5em minus 0.4em\relax IEEE, 2021, pp. 608--613.

\bibitem{ju2020deception}
Z.~Ju, H.~Zhang, and Y.~Tan, ``Deception attack detection and estimation for a local vehicle in vehicle platooning based on a modified ufir estimator,'' \emph{IEEE Internet of Things Journal}, vol.~7, no.~5, pp. 3693--3705, 2020.

\bibitem{zhang2020distributed}
D.~Zhang, Y.-P. Shen, S.-Q. Zhou, X.-W. Dong, and L.~Yu, ``Distributed secure platoon control of connected vehicles subject to dos attack: Theory and application,'' \emph{IEEE Transactions on Systems, Man, and Cybernetics: Systems}, vol.~51, no.~11, pp. 7269--7278, 2020.

\bibitem{mousavinejad2019distributed}
E.~Mousavinejad, F.~Yang, Q.-L. Han, X.~Ge, and L.~Vlacic, ``Distributed cyber attacks detection and recovery mechanism for vehicle platooning,'' \emph{IEEE Transactions on Intelligent Transportation Systems}, vol.~21, no.~9, pp. 3821--3834, 2019.

\bibitem{petrillo2020secure}
A.~Petrillo, A.~Pescape, and S.~Santini, ``A secure adaptive control for cooperative driving of autonomous connected vehicles in the presence of heterogeneous communication delays and cyberattacks,'' \emph{IEEE transactions on cybernetics}, vol.~51, no.~3, pp. 1134--1149, 2020.

\bibitem{dadras2015vehicular}
S.~Dadras, R.~M. Gerdes, and R.~Sharma, ``Vehicular platooning in an adversarial environment,'' in \emph{Proceedings of the 10th ACM Symposium on Information, Computer and Communications Security}, 2015, pp. 167--178.

\bibitem{dadras2019resilient}
S.~Dadras, S.~Dadras, and C.~Winstead, ``Resilient control design for vehicular platooning in an adversarial environment,'' in \emph{2019 American Control Conference (ACC)}.\hskip 1em plus 0.5em minus 0.4em\relax IEEE, 2019, pp. 533--538.

\bibitem{khanapuri2021learning}
E.~Khanapuri, T.~Chintalapati, R.~Sharma, and R.~Gerdes, ``Learning based vehicle platooning threat detection, identification and mitigation,'' 2021.

\bibitem{xu2019event}
W.~Xu, D.~W. Ho, J.~Zhong, and B.~Chen, ``Event/self-triggered control for leader-following consensus over unreliable network with dos attacks,'' \emph{IEEE transactions on neural networks and learning systems}, vol.~30, no.~10, pp. 3137--3149, 2019.

\bibitem{de2015input}
C.~De~Persis and P.~Tesi, ``Input-to-state stabilizing control under denial-of-service,'' \emph{IEEE Transactions on Automatic Control}, vol.~60, no.~11, pp. 2930--2944, 2015.

\bibitem{feng2019secure}
Z.~Feng and G.~Hu, ``Secure cooperative event-triggered control of linear multiagent systems under dos attacks,'' \emph{IEEE Transactions on Control Systems Technology}, vol.~28, no.~3, pp. 741--752, 2019.

\bibitem{he2021secure}
W.~He and Z.~Mo, ``Secure event-triggered consensus control of linear multiagent systems subject to sequential scaling attacks,'' \emph{IEEE Transactions on Cybernetics}, vol.~52, no.~10, pp. 10\,314--10\,327, 2021.

\bibitem{liang2021secure}
W.~Xu, G.~Hu, D.~W.~C.~Ho, and F.~Zhi, ``Distributed secure cooperative control under denial-of-service attacks from multiple adversaries,'' vol.~50, no.~8.\hskip 1em plus 0.5em minus 0.4em\relax IEEE, 2019, pp. 3458--3467.

\bibitem{xiao2020resilient}
S.~Xiao, X.~Ge, Q.-L. Han, Z.~Cao, Y.~Zhang, and H.~Wang, ``Resilient distributed event-triggered control of vehicle platooning under dos attacks,'' \emph{IFAC-PapersOnLine}, vol.~53, no.~2, pp. 1807--1812, 2020.

\bibitem{xu2022event}
Y.~Xu and G.~Guo, ``Event triggered control of connected vehicles under multiple cyber attacks,'' \emph{Information Sciences}, vol. 582, pp. 778--796, 2022.

\end{thebibliography}

\end{document}